\documentclass[a4paper,pre,reqno,superscriptaddress,twocolumn,showpacs]{revtex4-1}
\usepackage[T1]{fontenc}
\usepackage[latin9]{inputenc}
\usepackage{units}
\usepackage{mathrsfs}
\usepackage{mathtools}
\usepackage{amsmath}
\usepackage{multirow,enumitem}
\usepackage{amssymb}
\usepackage{graphicx}
\usepackage{esint}
\PassOptionsToPackage{normalem}{ulem}
\usepackage{ulem}
\usepackage{todonotes}
\usepackage{changes}
\usepackage{lipsum}
\usepackage{color}
\usepackage{array}

\newcommand{\eref}[1]{Eq.~(\ref{#1})}
\newcommand{\erefs}[1]{Eqs.~(\ref{#1})}

\def\bea{\begin{eqnarray}}
\def\eea{\end{eqnarray}}
\def\ba{\begin{array}}
\def\ea{\end{array}}
\def\n{\nonumber}

\def\la{\langle}
\def\ra{\rangle}

\usepackage{colortbl}
\usepackage{xcolor}
\usepackage{hyperref}

\usepackage{babel}

\begin{document}
\title{Harmonically trapped inertial run-and-tumble particle in one dimension}
\author{Debraj Dutta}
\affiliation{S. N. Bose National Centre for Basic Sciences, Kolkata 700106, India}
\author{Anupam Kundu}
\affiliation{International Centre for Theoretical Sciences, Bengaluru 560089, India}
\author{Sanjib Sabhapandit}
\affiliation{Raman Research Institute, Bengaluru 560080, India}
\author{Urna Basu}
\affiliation{S. N. Bose National Centre for Basic Sciences, Kolkata 700106, India}

\begin{abstract}
We study the nonequilibrium stationary state of a one-dimensional inertial run-and-tumble particle (IRTP) trapped in a harmonic potential.  We find that the presence of inertia leads to two distinct dynamical scenarios, namely, overdamped and underdamped, characterized by the relative strength of the viscous and the trap time-scales. We also find that inertial nature of the active dynamics leads to the particle being confined in specific regions of the phase plane in the overdamped and underdamped cases, which we compute analytically. Moreover, the interplay of the inertial and active time-scales gives rise to several sub-regimes, which are characterized  by very different behaviour of position and velocity fluctuations of the IRTP. In particular, in the underdamped regime, both the position and velocity undergoes transitions from a novel multi-peaked structure in the strongly active limit to a single peaked Gaussian-like distribution in the passive limit. On the other hand, in the overdamped scenario, the position distribution shows a transition from a U-shape to a dome-shape, as activity is decreased. Interestingly, the velocity distribution in the overdamped scenario shows two transitions---from a single-peaked shape with an algebraic divergence at the origin in the strongly active regime to a double peaked one in the moderately active regime to a dome-shaped one in the passive regime. 
\end{abstract}

\maketitle

\section{Introduction}

Active particles perform self-propelled motion by consuming energy at an individual level and thereby breaking detailed balance. Active motion is observed in nature over multiple length scales ranging from bacterial motility, cell migration, and insect movements, to bird flocks and fish schools at macroscopic scales
\cite{cavagna2014bird, feinerman2018physics, pavlov2000patterns, mukundarajan2016surface}. Recent years have also seen a huge advancement in the synthesis of artificial active agents such as nanobots, Janus colloids to granular particles and self-propelled robots \cite{yang2012janus, Bechinger2016, walther2013janus, nelson2010microrobots}. Most commonly, the motions of individual active agents are described theoretically using overdamped Langevin equations, such as Active Brownian process (ABP)~\cite{basu2018active, howse2007self}, Run-and-Tumble process (RTP)~\cite{Malakar2018, tailleur2008statistical} and Active Ornstein-Uhlenbeck process (AOUP)~\cite{bonilla2019active, martin2021statistical}. In such a description, the inertial effects are ignored assuming that the velocity relaxes over a time-scale much smaller than the observation time-scale. 

The typical relaxation time of the velocity of an object of linear size $\ell$ and density $\rho$ scales as $\rho \ell^2$. For example, for a micron-sized particle, the typical relaxation time of the velocity is around $100$~ns, while the typical temporal resolution of experiments is usually above microseconds, making the overdamped Langevin description fairly accurate. On the other hand, for a macro-sized particle, say, with $\ell \sim 1$~mm,  the relaxation time is about $0.1$~s, necessitating the inclusion of inertia in the description of its motion.  Therefore, to describe the motion of macro-sized active particles, one needs to consider the underdamped Langevin equations.

Recent years have seen an increased effort to explore the effect of inertia on active particle motion, both at individual and collective levels. It was found that the presence of inertia drastically changes the collective properties of active matter. For example, it changes the nature of motility induced phase transition~\cite{mandal2019motility, su2021inertia}, hinders crystallization~\cite{de2020phase, liao2021inertial}, promotes hexatic ordering in homogeneous phases~\cite{negro2022hydrodynamic} and, in general, reduces spatial velocity correlations~\cite{marconi2021hydrodynamics, caprini2021collective, caprini2021spatial}. It has also been shown that the presence of inertia makes an active reservoir behave more like an equilibrium one~\cite{khali2024active}. 

The inclusion of inertia also leads to remarkable effects on the statistical properties of individual active particles. It has been shown that adding inertia to ABPs impacts their behaviour both in free space as well as in the presence of external potential, manifest in the change of kinetic temperature and swim pressure~\cite{patel2023exact, patel2024exact, sprenger2023dynamics}. Moreover, presence of rotational inertia also affects the long-time diffusivity for an ABP \cite{scholz2018inertial, breoni2020active,lisin2022motion}. The effect of inertia on the large deviation function of a harmonically trapped RTP in the strongly passive regime has recently been studied~\cite{smith2022nonequilibrium}.  The dynamics of inertial AOUP has also been investigated recently both for free particles and harmonically trapped particles~\cite{caprini2021inertial}, as well as in the presence of underdamped self-propulsion force~\cite{sprenger2023dynamics}. The behaviour of inertial ABP and AOUP have also been studied in the presence of a magnetic field~\cite{muhsin2022inertial, obreque2024dynamics}, coulomb friction~\cite{antonov2024inertial} and in viscoelastic media~\cite{muhsin2023inertial, adersh2024inertial}. Since AOUP is a linear Gaussian process, the corresponding position and velocity distributions are Gaussian, completely characterized by the correlation matrix.  For non-Gaussian processes like ABP and RTP, a comprehensive understanding of the effect of inertia on the full position and velocity distributions is still lacking. 

In this paper, we address this question in the context of an inertial run-and-tumble particle in the presence of a harmonic potential. Run-and-tumble particle, which is one of the simplest and most-studied model of active motion, describes the overdamped motion of a particle along an internal direction which itself changes stochastically \cite{Malakar2018, tailleur2008statistical}. Inertia induces a longer delay for the particles to change their speed, increasing the persistence of trajectories and particle mobility. In the presence of a trap, the particle reaches a nonequilibrium stationary state. We find that the interplay between inertia and self-propulsion leads to non-trivial features in the stationary position and velocity distributions, that we characterize analytically. 

The paper is organized as follows. In the next section we introduce the the inertial Run-and-Tumble particle (IRTP) model and provide a summary of our main results. We then investigate the behaviour of the position and velocity moments in Sec.~\ref{sec:moments} to get an idea about the typical position and velocity fluctuations. In Sec.~\ref{ud_case} and Sec.~\ref{sec:OD}, we provide a detailed discussion of the stationary position and velocity distributions in the context of different regimes and sub-regimes arising from the interplay of different time-scales. Finally, we conclude in Sec.~\ref{conc} with some open questions. 

\begin{figure}[t]
\begin{centering}
\includegraphics[width=8cm]{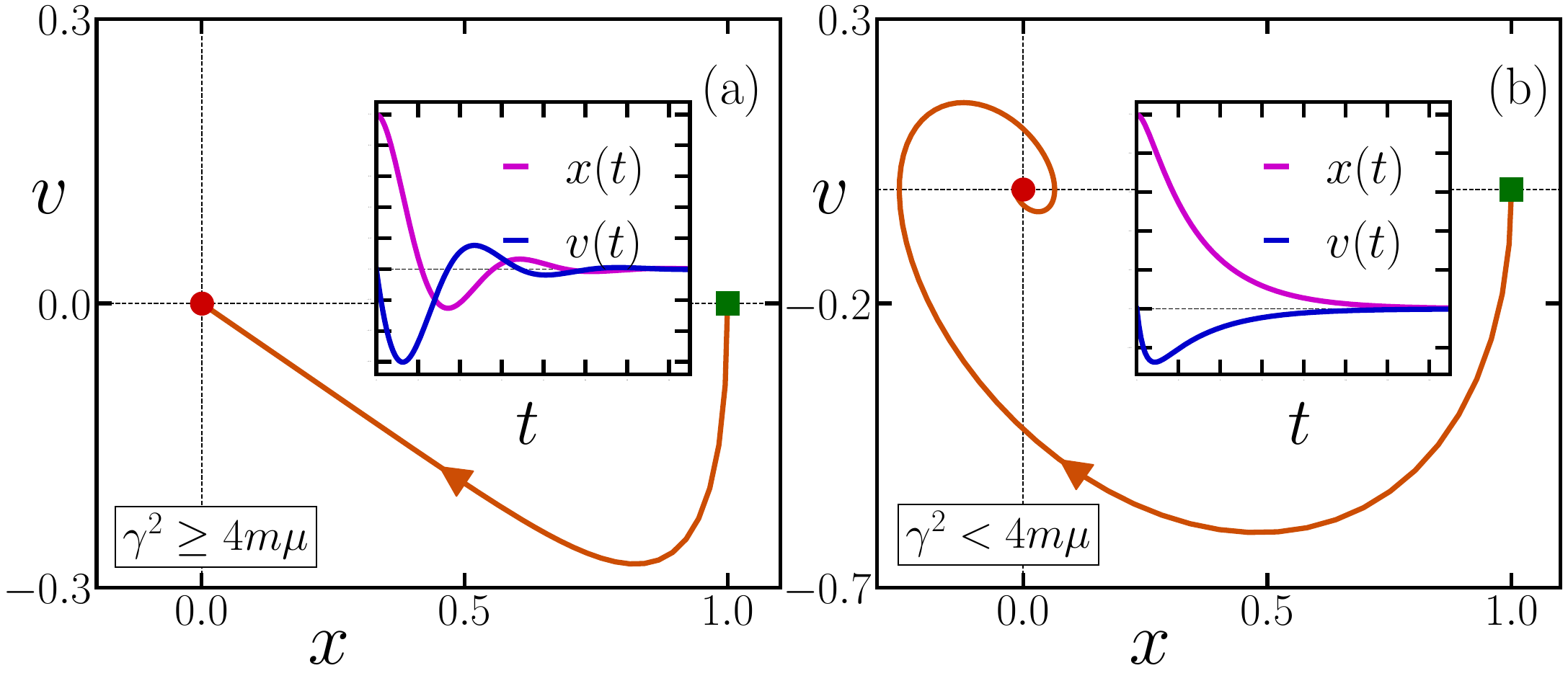}
\end{centering}
\caption{Phase space trajectory of a damped harmonic oscillator for the overdamped (a) and underdamped (b) scenarios, starting from $(x_0,v_0)=(1,0)$ [indicated by dark green squares]. The insets show the corresponding time-evolution of $x(t)$ and $v(t)$ separately. In both the cases, the particle eventually reaches $(0,0)$ [indicated by red discs]. Here we have used $\mu=1.0,m=1.0,a_0=1.0$ with $\gamma=3.0$ for the overdamped case and $\gamma=0.8$ for the underdamped case.}
\label{fig:traj_det_odud}
\end{figure}

\section{Model and summary of the results}
We consider the one-dimensional motion of an inertial run-and-tumble particle (IRTP) in a harmonic potential $U(x)=\mu x^2 / 2$. The position $x(t)$ and the velocity $v(t)$ of the particle evolve according to the Langevin equations,
\begin{align}
\dot{x}(t) & = v(t), \label{eq:Leq1}\\
m\dot{v}(t) & = - \mu x(t) - \gamma v(t) + a_{0}\sigma(t), \label{eq:Leq2}
\end{align}
where $\gamma$ denotes the dissipation coefficient. The activity of the particle originates from the self-propulsion force $a_0 \sigma(t)$, where $a_0$ is a constant and $\sigma(t)$ is a dichotomous noise, alternating between $\pm 1$ with a constant rate $\tau_a^{-1}$. Note that, between two such successive `tumbling' events, i.e., when $\sigma$ remains constant, the particle undergoes a deterministic damped motion in a harmonic trap centered at $a_0\sigma/\mu$. The tumbling results in a switch in the position of the trap between $\pm a_0/\mu$. 

In general, there can also be an additional Gaussian white noise with a strength $\sqrt{2\gamma k_{B}T}$ in Eq.~\eqref{eq:Leq2}, arising from thermal fluctuations, which can be neglected in most physical situations. Moreover, mathematically, the effect of the white noise can be incorporated straightforwardly as a convolution with the $T=0$ solution. Hence, it suffices to consider the $T=0$ case. We are interested in the position and velocity fluctuations of this IRTP. In particular, we investigate how the presence of the finite mass $m$ affects the marginal position and velocity distributions. 

\begin{figure}[t]
\begin{centering}
\includegraphics[width=8cm]{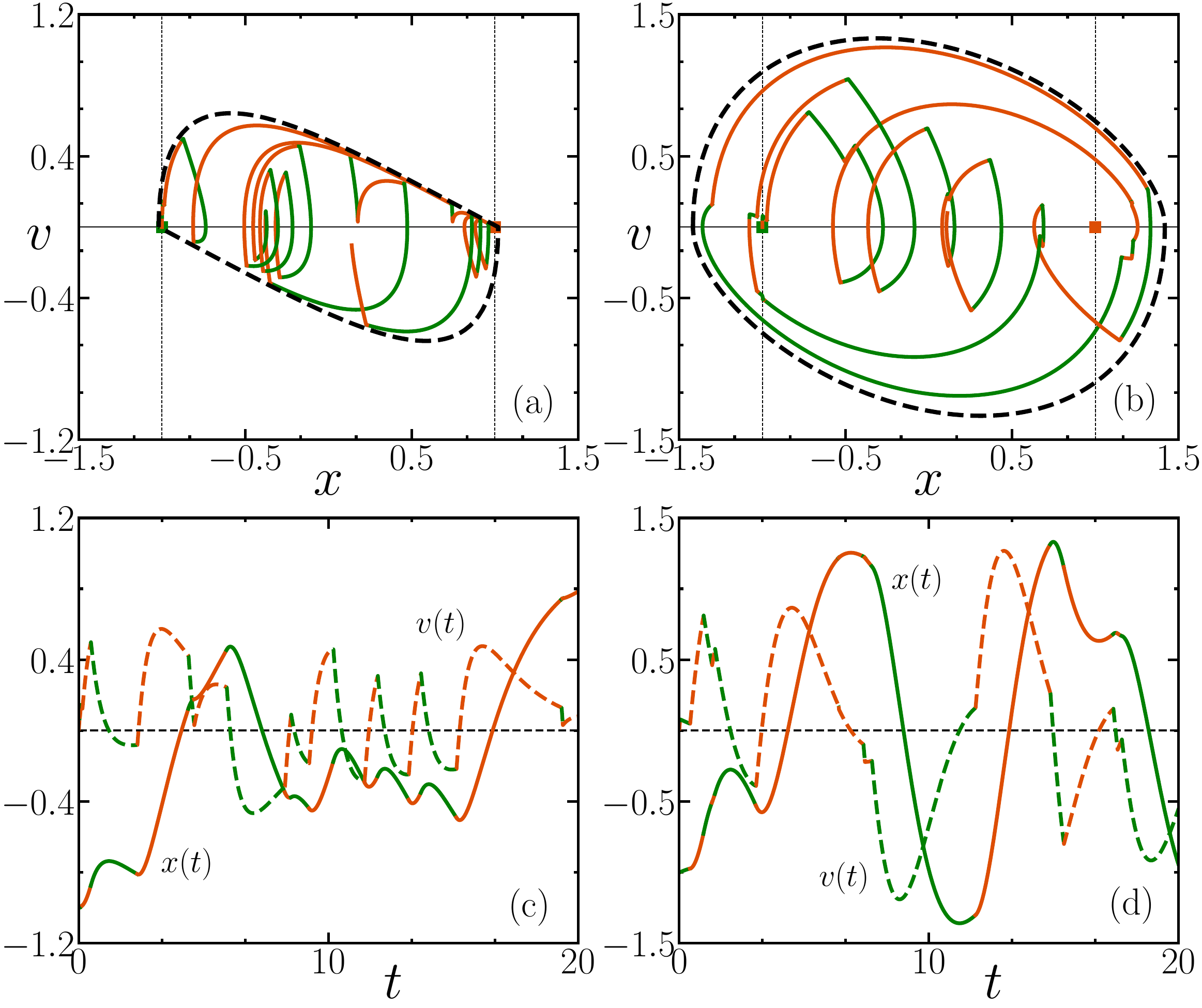}
\end{centering}
\caption{Typical phase trajectory for the IRTP for the overdamped (a) and underdamped (b) scenarios. The orange and green solid lines correspond to trajectories with $\sigma=\pm 1$ respectively and the dotted vertical lines indicate the alternating positions $\pm a_0/\mu$ of the trap minima. Additionally, the dashed vertical line marks the boundary of the support of the joint distribution $P_{\text{st}}(x,v)$ in the phase plane. The corresponding time evolution of $x(t)$ and $v(t)$ are also shown separately for the overdamped (c) and underdamped (d) cases. Here we have used $\tau_a=1,\mu=1,m=1$ with $\gamma=1$ for the overdamped case and $\gamma=2.5$ for the underdamped case.}
\label{fig:traj_oud}
\end{figure}

In the absence of the active noise, i.e., for $a_0=0$, equations \erefs{eq:Leq1}-\eqref{eq:Leq2} describe the deterministic motion of a particle in a harmonic trap centered at the origin, where the particle eventually  comes to rest at the minimum of the potential. In the $\mu \to 0$ limit, the viscous time-scale $\tau_1=m/\gamma$ controls the relaxation of the velocity. On the other hand, when $m \to 0$, the time-scale $\tau_2=\gamma/\mu$ controls the relaxation of the particle position in the harmonic trap. For non-zero $m$ and $\mu$, the presence of the two time-scales $\tau_1$ and $\tau_2$ give rise to two scenarios, namely,
\begin{itemize}
    \item Underdamped case: $\gamma^2 < 4m\mu$, i.e., $\tau_2<4\tau_1$, 
    \item Overdamped case: $\gamma^2 \ge 4m\mu$, i.e., $\tau_2\ge4\tau_1$,
\end{itemize}
where the particle exhibits qualitatively different kinds of motion [see Fig.~\ref{fig:traj_det_odud}]. In the presence of the stochastic active force, the particle shows a rather interesting random motion [see Fig.~\ref{fig:traj_oud} for typical trajectories].  In fact, in this case, an additional time-scale $\tau_a$ appears, which characterize the activity of the system. The relative strength of $\tau_a$ with respect to $\tau_1$ and $\tau_2$ and $\tau_a$ gives rise to a range of sub-regimes for the IRTP dynamics, in both the overdamped and underdamped scenarios. In this paper we characterize the position and velocity fluctuations of the IRTP in all these regimes.

\begin{figure*}[t]
\begin{centering}
\includegraphics[width=17 cm]{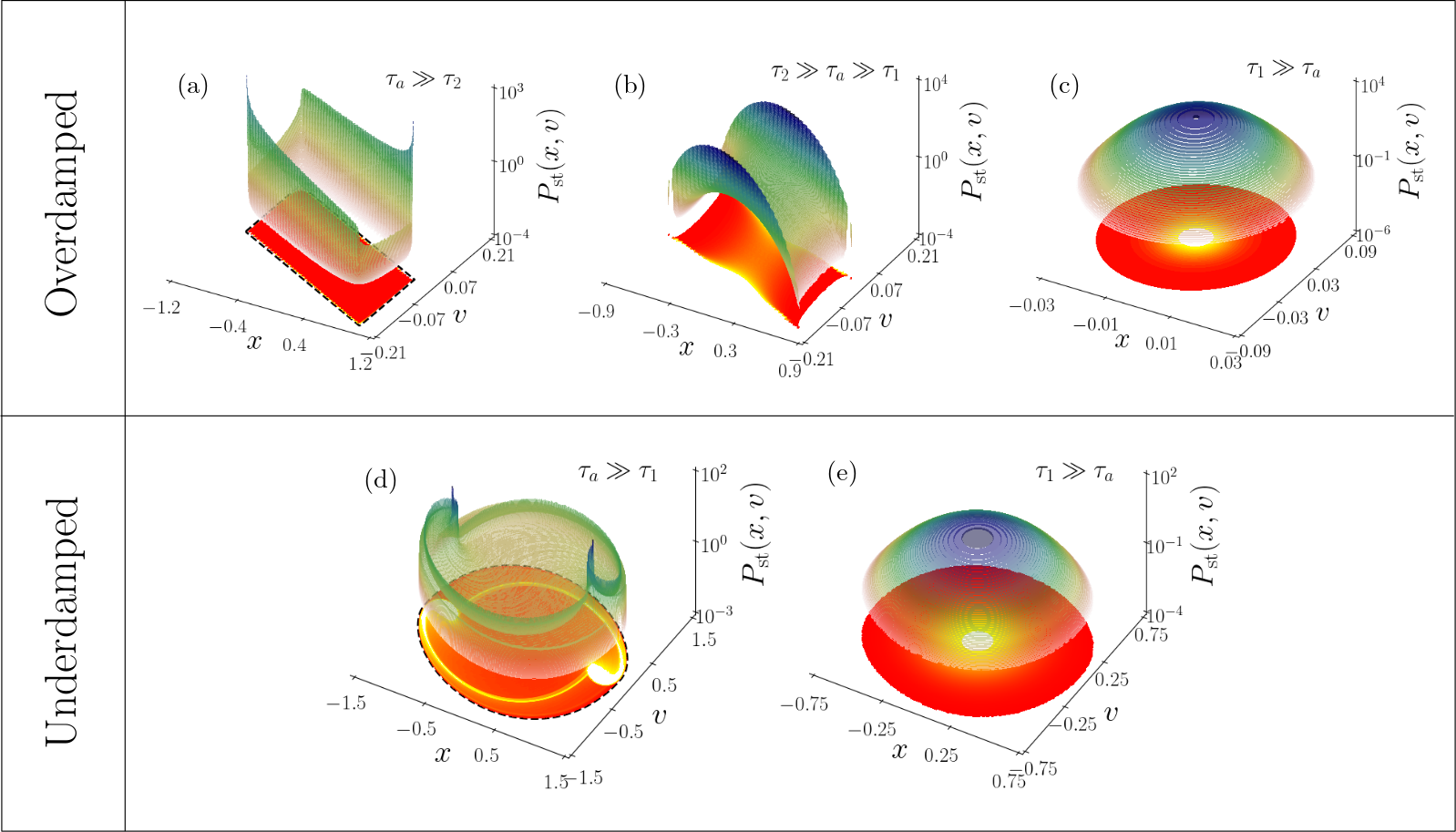}
\end{centering}
\caption{Plot for the joint stationary distribution $P_{\text{st}}(x,v)$ in the strongly overdamped ($\tau_1\ll\tau_2$) [(a), (b), (c)] and underdamped ($\tau_2\ll\tau_1$) [(d), (e)] scenarios, obtained using numerical simulations. In both cases, the distributions show a shape transition with changing activity (from left to right). The overdamped case is characterized by three qualitatively distinct shapes of the joint distribution $P_\text{st}(x,v)$ in three different activity regimes. In contrast, the underdamped case is characterized by two activity regimes, resulting in two qualitatively different behaviours of $P_\text{st}(x,v)$. The analytically obtained bound for the joint distribution $P_{\text{st}}(x,v)$ in the (a) overdamped [see \erefs{eq:lcyc_3}-\eqref{eq:lcyc_4}] and (d) underdamped [see \erefs{eq:lcyc_1}-\eqref{eq:lcyc_2}] cases are also drawn on the $x-v$ plane with black-dashed line. Here, we have used parameter values $m=0.1,\mu=1.0,a_0=1.0,\gamma=10.0$ and $\tau_a=100.0, 1.0, 0.001$ (left to right) for the overdamped case and $m=1.0,\mu=1.0,\gamma=1.0,a_0=1.0$ and $\tau_a=2.5, 0.1$ (left to right) for the underdamped case.}\label{fig:Pxv_ud_od}
\end{figure*}

Note that, in the viscous limit $\tau_1 \to 0$, \eref{eq:Leq2} reduces to the well-studied ordinary run-and-tumble particle (RTP) \cite{Malakar2018} with self-propulsion speed $\bar{v}_0$ in a harmonic trap of strength $\bar{\mu}$, where,
\begin{align}
\bar{v}_0=a_0/\gamma, ~~\text{and} \quad \bar{\mu}=\mu/\gamma. \label{eq:v0_mu_def} 
\end{align}
In this case, the stationary state position distribution, bounded in the regime $ -\bar{v}_0/\bar \mu \le x \le \bar v_0/\bar \mu$, undergoes a shape transition at a critical activity $\tau_{a}^{\star} = \gamma / \mu$ \cite{dhar2019run}.  The presence of inertia is expected to affect both position and velocity fluctuations, which we investigate here. 

\newcolumntype{P}[1]{>{\raggedright\arraybackslash}p{#1}}

\begin{table*}[]
\renewcommand{\arraystretch}{1.5}
\centering
\begin{tabular}{|P{3 cm}|| P{4 cm}|P{4 cm}|P{4 cm}|}
  \hline
  {\bf Strongly Overdamped} ($\tau_1 \ll \tau_2$): \newline 

  \begin{itemize}[leftmargin=*]
      \item   Position bound $|x| \le  X_\text{OD}$  \newline [see \eref{eq:x_b_}] 

      \item  Velocity bound $|v| \le V_\text{OD}$ \newline [see \eref{eq:v_b_} ]
  \end{itemize}

  & {\bf Strongly active regime} ($\tau_2 \ll \tau_a$)~[see Fig.~\ref{fig:Pxv_ud_od}(a)]
    \newline
    \begin{itemize}[leftmargin=*]
        \item[$\circ$] Position distribution: U-shaped with divergences at $x = \pm X_{\text{OD}}$.
        \item[$\circ$] Velocity distribution: Single algebraic divergence at $v=0$.
    \end{itemize}

  & {\bf Moderately active regime} ($\tau_1 \ll \tau_a \ll \tau_2$)~[see Fig.~\ref{fig:Pxv_ud_od}(b)]
    \newline
    \begin{itemize}[leftmargin=*]
        \item[$\circ$] Position distribution: Dome shaped with peak at $x=0$.
        \item[$\circ$] Velocity distribution: Double peaked with minimum at $v=0$.
    \end{itemize}

  & {\bf Strongly passive regime} ($\tau_a \ll \tau_1$)~[see Fig.~\ref{fig:Pxv_ud_od}(c)]
    \newline
    \begin{itemize} [leftmargin=*]
        \item[$\circ$] Position distribution: Gaussian around $x=0$   
        \item[$\circ$] Velocity distribution: Gaussian around $v=0$
    \end{itemize} 
  
    \\ \hline
  
  {\bf Strongly Underdamped} ($\tau_2 \ll \tau_1$):
  \newline 

  \begin{itemize}[leftmargin=*]
      \item   Position bound $|x| \le  X_\text{UD}$  \newline [see \eref{eq:x_b}] 

      \item  Velocity bound $|v| \le V_\text{UD}$ \newline [see \eref{eq:v_b}]
  \end{itemize}
  
  & \multicolumn{2}{P{8 cm}|}{{\bf Strongly active regime }($\tau_1 \ll \tau_a$)~[see Fig.~\ref{fig:Pxv_ud_od}(d)] 
        \newline
        \begin{itemize}[leftmargin=*]
            \item[$\circ$] Position distribution: Multi-peaked with preliminary peaks at $x=\pm a_0/\mu$.
            \item[$\circ$] Velocity distribution: Multi-peaked with preliminary peak at $v=0$.
        \end{itemize}} 

  & {\bf Strongly passive regime} ($\tau_a \ll \tau_1$)~[see Fig.~\ref{fig:Pxv_ud_od}(e)] 
    \newline
    \begin{itemize} [leftmargin=*]
        \item[$\circ$] Position distribution: Gaussian around $x=0$   
        \item[$\circ$] Velocity distribution: Gaussian around $v=0$
    \end{itemize} 
    
    \\ \hline
    
\end{tabular}
\caption{Summary of the different regimes emerging from the interplay of the inertial time-scale $\tau_1=m/\gamma$, the viscous time-scale $\tau_2=\gamma/\mu$ and the activity time-scale $\tau_a$.
}
\label{tab:summary}
\end{table*}

We first discuss the overdamped case ($4\tau_1\le\tau_2$). The linear nature of the Langevin equations~\eqref{eq:Leq2} allows us to solve it exactly, albeit formally [see Appendix \ref{ap:solution_Langevin} for the details]. In the overdamped scenario, the formal solution reads,
\begin{align}
x(t) =& \frac{2\bar{v}_{0}}{\lambda }\int_{0}^{t}ds~\sigma(t-s)\exp\Big[-\frac {s}{2\tau_1}\Big]\sinh\frac{\lambda  s}{2\tau_1}, \label{eq:xt_od} \\
v(t) =& \frac{\bar{v}_0}{\tau_1}\int_{0}^{t}ds\,\sigma(t-s)\exp\Big[-\frac{s}{2\tau_1}\Big]\Big\{\cosh\frac{\lambda  s}{2\tau_1} \cr
 & \qquad \qquad \qquad \qquad - \frac{1}{\lambda }\sinh\frac{\lambda  s}{2\tau_1}\Big\}, \label{eq:vt_od}
\end{align}
where we have defined, 
\bea
\lambda =\sqrt{1 - \frac {4\tau_1}{\tau_2}}\quad\text{with}\quad\tau_1 = \frac{m}{\gamma}, ~\tau_2=\frac{\gamma}{\mu}.\label{eq:w_def}
\eea
Here we assume, without any loss of generality, that at $t=0$ the particle is at rest at the origin, i.e., $x(0)=0,v(0)=0$. Evidently, in this overdamped regime, the combination of the two time-scales $\tau_1$ and $\tau_2$ gives rise to the emergent time-scales $\tau_1/(1\pm \lambda )$. However, for the strongly overdamped scenario $\tau_1 \ll \tau_2$, these time-scales reduce to the intrinsic ones $\tau_1$ and $\tau_2$, since in this limit, $1 + \lambda  \simeq 2$ and $1 - \lambda  \simeq 2\tau_1/\tau_2$. The interplay of the three time-scales, $\tau_1,\tau_2$ and ${\tau_a}$ leads to the emergence of three regimes of activity, namely, (i) the strongly active regime ($ \tau_1, \tau_2\ll\tau_a $), (ii) the moderately active regime,($\tau_1\ll\tau_a\ll\tau_2$) and (iii) the passive regime $\tau_a\ll\tau_1,\tau_2$ [see Table \ref{tab:summary}]. We analyse the behaviour of the position and velocity distributions in these regimes in Sec \ref{sec:OD}. 

Next, we discuss the underdamped oscillatory scenario ($\tau_2< 4\tau_1$), where the solution of \eref{eq:Leq2} is given by,
\begin{align}
x(t) =& \frac{2\bar{v}_{0}}{\Lambda}\int_{0}^{t}ds~\sigma(t-s)\exp\Big[-\frac{s}{2\tau_1}\Big]\sin\frac{\Lambda s}{2\tau_1}, \label{eq:xt_ud} \\
v(t) =& \frac{\bar{v}_{0}}{\tau_1}\int_{0}^{t}ds\,\sigma(t-s)\exp\Big[-\frac{s}{2\tau_1}\Big]\Big\{\cos\frac{\Lambda s}{2\tau_1} \cr
 & \qquad \qquad \qquad \qquad - \frac{1}{\Lambda}\sin\frac{\Lambda s}{2\tau_1}\Big\}, \label{eq:vt_ud}
\end{align}
with, 
\bea
\Lambda=\sqrt{\frac{4 \tau_1}{\tau_2} - 1}. \label{eq:Lamb_def}
\eea
Here, once again, we have assumed $x(0)=0$ and $v(0)=0$.

Clearly, from Eqs.~\eqref{eq:xt_ud}-\eqref{eq:vt_ud}, the time-scale $\tau_2$, appearing through $\Lambda$, only governs the oscillation period, and the relaxation is solely determined by the time-scale $\tau_1$. Therefore, the interplay of this time-scale $\tau_1$ and the active time-scale $\tau_a$ gives rise to only two regimes $\tau_1\ll\tau_a$ (strongly active) and $\tau_a\ll\tau_1$ (strongly passive) [see Table \ref{tab:summary}].

The bounded nature of the active  noise $\sigma(t)$ ensures that the stationary distribution $P_{\text{st}}(x,v)$ is supported in a finite region of the $x-v$ phase plane. However, the supports are different in the overdamped and underdamped cases. Figure~\ref{fig:traj_oud}(a) shows a typical trajectory of the IRTP in the overdamped scenario which indicates that the particle position, in this case, is always confined between the two possible trap centres $\pm a_0/\mu$. In fact, we  compute the boundary of the supporting region in the phase space [see Fig.~\ref{fig:bound_od} and \erefs{eq:lcyc_3}-\eqref{eq:lcyc_4}]. In particular, we show that the marginal position and velocity distributions are supported in the regions,
\begin{align}
|x| \leq X_\text{OD} &\equiv \frac{a_0}{\mu},\label{eq:x_b_}\\
|v| \leq V_\text{OD} &\equiv \frac{2 a_0}{\sqrt{m \mu}}\exp\Big[-\frac{1}{\lambda }\text{sech}^{-1}\frac{2\sqrt{m\mu}}{\gamma}\Big]\label{eq:v_b_},
\end{align}
respectively, where $\lambda $ is defined in \eref{eq:w_def}. On the other hand, in the underdamped case, the inertia allows the particle to move past the trap centres, thereby extending the accessible region in the phase space [see Fig.~\ref{fig:traj_oud}(b)]. We show that, in this case, the bound
depends on the mass of the particle [see Fig.~\ref{fig:bound_traj_ud} and \erefs{eq:lcyc_1}-\eqref{eq:lcyc_2}]. In particular, we find that the marginal distributions are supported in the region,
\begin{align}
|x| \le  X_{\text{UD}} &\equiv \frac{a_{0}}{\mu}\coth \frac{\pi}{2\Lambda},\label{eq:x_b}\\
|v| \le  V_{\text{UD}} &\equiv \frac{a_0}{\sqrt{m\mu}}\Big[1+\coth\frac{\pi}{2\Lambda}\Big]\exp\bigg[-\frac{1}{\Lambda}\tan^{-1}\Lambda\bigg],\label{eq:v_b}
\end{align}
where $\Lambda$ is defined in \eref{eq:Lamb_def}.  Figure~\ref{fig:traj_oud} illustrates typical trajectories of the particle on the $x-v$ plane as well as time evolution  of $x(t)$ and $v(t)$, in both the scenarios.

The stationary distribution $P_{\text{st}}(x,v)$ in the underdamped and overdamped scenarios, obtained from numerical simulations, are shown in Fig.~\ref{fig:Pxv_ud_od}. In both the cases, the distributions undergo intricate shape transitions, from a multi-peaked structure in the strongly active regime to a single-peaked one in the passive regime.  However, the multi-peaked structures appearing in the strongly active regime are very different in the overdamped and the underdamped scenarios. To analytically characterize these shape transitions, we focus on the marginal position and velocity distributions separately. 

In the strongly active regime, the distributions can be obtained by considering the deterministic motion of the particles between successive tumbling events. Nevertheless, the difference in the nature of the deterministic trajectories in the overdamped and underdamped cases leads to drastically different shapes of the distributions. In the former case, the peaks in the marginal position distribution always appear at the edges of the support $\pm X_\text{OD}$ whereas in the latter case, multiple peaks appear inside the support. On the other hand, the marginal distribution of velocity in the overdamped case has a single peak at $v=0$ in contrast to a multiple peaked structure in the underdamped case. These peaks smoothen out as the activity is decreased and eventually, in the passive limit, both the position and velocity distributions converge to the Boltzmann measure $P_\text{eq}(x,v) \propto e^{-\beta_\text{eff}(U(x) + m v^2/2)}$ with $\beta_\text{eff} = 2\gamma/(\tau_a a_0^{2})$.

For the overdamped case, the transition from the strongly active to the passive regime goes through an intermediate, moderately active, regime as shown in Fig.~\ref{fig:Pxv_ud_od} (a)-(c) [also see Table \ref{tab:summary}]. In particular, we compute the position and velocity distributions analytically in the strongly overdamped limit ($\gamma^2\gg m\mu$). The position distribution undergoes a transition from a U-shape in the strongly active regime to a dome-shape in the moderately active regime at a critical  activity, $\tau_a^{\star}(m)\equiv 1/\kappa=\tau_a^{\star}[1- 2 \tau_1 /\tau_2 + O(\tau_1^2)]$, where $\kappa$ is defined in \eref{eq:kappa_nu_def}. Eventually, in the strongly passive regime, the position distribution converges to the Gaussian. On the other hand, we find the velocity distribution in the active regime as,
\begin{align}
P_{\text{st}}(v)=\frac{2^{\alpha}}{\nu B(\alpha,\alpha)}\Big(\frac{|v|}{\nu}\Big)^{\alpha-1}\Big(1-\frac{|v|}{2\nu}\Big)^{\alpha}, 
\text{with} ~~ \alpha = \frac 1{\tau_a\kappa},  \label{eq:Pv_eff}
\end{align}
where $B(u,u)$ denotes the beta function~\cite[\href{https://dlmf.nist.gov/5.12.E1}{(5.12.1)}]{NIST:DLMF} and $\nu$ is defined in \eref{eq:kappa_nu_def}. Clearly, the velocity distribution also undergoes a shape transition at $\alpha=1$, i.e., $\tau_{a}^{\star}(m)=1/\kappa$, from a distribution exhibiting algebraic divergence at $v=0$ with exponent $1-1/(\tau_a\kappa)$, in the strongly active regime $\tau_a^{\star}(m) \ll \tau_a$, to a bimodal distribution with a minimum at $v=0$ in the moderately active regime. Interestingly, the velocity distribution undergoes an additional shape transition to a dome-shaped distribution, as the activity is decreased further, eventually becoming Gaussian in the strongly passive regime [see Fig.~\ref{fig:od_approx_v}]. Table~\ref{tab:summary} briefly summarises the results in the different regimes and subregimes.

In the following sections, we analyze the position and velocity distributions of the IRTP in the overdamped and underdamped regimes separately. We start with computing the moments and correlations of position and velocity in the next section.

\section{Moments}\label{sec:moments}

To get an idea about the nature of the position and velocity fluctuations, we first investigate the behaviour of the moments of $x$ and $v$ in the stationary state. Let  $P_\sigma(x,v,t)$ denote the joint probability density for the particle to have position $x$ and velocity $v$, along with self-propulsion force direction $\sigma$, at time $t$. It is straightforward to see that the set of Fokker-Planck equations describing the evolution of $P_{\sigma}(x,v,t)$ are given by,
\bea
\frac{\partial P_{\sigma}}{\partial t} &=& - \frac{\partial}{\partial x}(vP_{\sigma}) +  \frac 1m \frac{\partial}{\partial v}\big[(\gamma v+\mu x-a_{0}\sigma ) P_\sigma\big] \n \\ [0.25em]
 && -(P_\sigma - P_{-\sigma})/\tau_a. \label{eq:Fp_eqns}
\eea 
It is difficult to solve this Fokker-Planck equation directly, even in the stationary state. However, as we show below, the stationary correlations of the form $\la x^{k} v^{n} \ra$, for any arbitrary values of $k$ and $n$, can be computed recursively. 

To this end, we define the quantities $P(x,v,t)=\sum_{\sigma}P_{\sigma}(x,v,t)$ and  $Q(x,v,t)=\sum_{\sigma}\sigma P_{\sigma}(x,v,t)$. Recasting the Fokker Planck equations \eqref{eq:Fp_eqns} in terms of $P$ and $Q$, and taking the stationary limit, $\partial_{t}P= 0$ and $\partial_{t}Q= 0$ we get,
\begin{align}
\frac{a_{0}}{m}\frac{\partial Q}{\partial v} & =-v\frac{\partial P}{\partial x}+\frac{\gamma}{m}\frac{\partial}{\partial v}(vP)+\frac{\mu}{m} x\frac{\partial P}{\partial v},\label{eq:fp3+} \\
\frac{a_{0}}{m}\frac{\partial P}{\partial v} & =-v\frac{\partial Q}{\partial x}+\frac{\gamma}{m}\frac{\partial}{\partial v}(vQ)+\frac{\mu}{m} x\frac{\partial Q}{\partial v}-\frac{2}{\tau_a}Q.\label{eq:fp3-}
\end{align}
We further define the correlation functions,
\bea 
M(k,n) &=& \intop_{-\infty}^{\infty}dx\intop_{-\infty}^{\infty}dv \, x^{k}v^{n}P(x,v),\label{eq: ML_eq1}\\
L(k,n) &=& \intop_{-\infty}^{\infty}dx\intop_{-\infty}^{\infty}dv \, x^{k}v^{n}Q(x,v),\label{eq: ML_eq2}
\eea 
where $k$ and $n$ are non-negative integers. Clearly, $M(k,n) \equiv \left\langle x^{k}v^{n}\right\rangle $ refer to the correlations we are interested in. Multiplying both sides of Eqs.~(\ref{eq:fp3+}) and (\ref{eq:fp3-}) with $x^{k}v^{n}$, and integrating over $x$ and $v$, we get a set of coupled recursion relations,
\begin{align}
&\mu n M(k+1,n-1) = mkM(k-1,n+1) -\gamma n M(k,n)\cr
& + a_{0}nL(k,n-1),~~\label{eq:rec1M} \\ 
&\mu n L(k+1,n-1) = mkL(k-1,n+1) -\gamma n L(k,n) \cr
& + a_{0}nM(k,n-1)-2mL(k,n)/\tau_a.~~\label{eq:rec1L}
\end{align}

Note that, to get the above equations we have used the fact that both $P(x,v)$ and $Q(x,v)$ vanish at $x \to \pm \infty$ and $v \to \pm \infty$. 
 
By definition, $M(k,n)=0$ for negative values of $k<0$ and $n<0$ and normalization of $P(x,v)$ gives the boundary condition $M(0,0)=1$. We further note that the \erefs{eq:fp3+}-\eqref{eq:fp3-} are invariant under the transformation $x\rightarrow-x,\,v\rightarrow-v,\,\sigma\rightarrow-\sigma$, which ensures,
\begin{align}
M(k,n)=0, & \qquad\forall(k+n)\in\text{odd},\\
L(k,n)=0, & \qquad\forall(k+n)\in\text{even}.\label{eq:constr1}
\end{align}
Substituting $n=0$ in \erefs{eq:rec1M}-\eqref{eq:rec1L}, we obtain two more constraints on $M(k,n)$ and $L(k,n)$, 
\begin{equation}
M(k,1)=0,\quad L(k,0)=\frac{k\tau_a}{2}L(k-1,1)\quad \text{for}~ k \ge 0.\label{eq:constr2}
\end{equation}
The relations \erefs{eq:rec1M}-\eqref{eq:constr2}, along with the normalization condition, $M(0,0)=1$, allow the computations of the correlations $M(k,n)$ and $L(k,n)$ recursively, as discussed below. 

\begin{figure}
\begin{centering}
\includegraphics[width=6.cm]{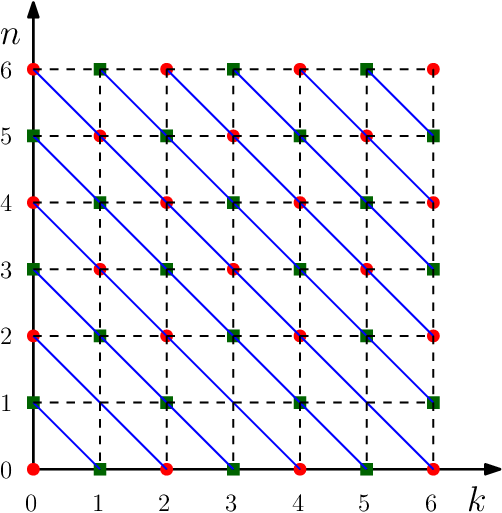}
\end{centering}
\caption{Schematic diagram illustrating the recursive connections between the correlation function : $M(k,n)$ (red disks) and $L(k,n)$ (green squares) drawn on the $k-n$ lattice.}
\label{fig: ML_scheme1}
\end{figure}

Figure  \ref{fig: ML_scheme1} shows the $k-n$ lattice where the $d$-th diagonal, given by $k+n=d$, contains $L(k,n)$ or $M(k,n)$ for odd and even $d$, respectively. We start with the first diagonal $k+n=1$ which contains the entries $L(1,0)$ and $L(0,1)$. These quantities satisfy 
\begin{align}
\mu L(1,0)=-\gamma L(0,1)+a_{0}M(0,0)-2 m L(0,1)/\tau_a,
\end{align}
which is obtained by setting $k=1,n=0$ in \eref{eq:rec1L}, along with the boundary condition $M(0,0)=1$. Moreover, we have, from \eref{eq:constr2} 
\begin{align}
\mu L(1,0)=\frac {\tau_a}{2}L(0,1).
\end{align}
The above equations can be solved immediately to obtain, 
\bea 
L(0,1) &=& \frac{2 a_{0} \tau_a }{(\mu\tau_a^2 + 2\gamma \tau_a +4 m)}, \cr
L(1,0) &=& \frac{ a_{0}\tau_a^2 }{(\mu\tau_a^2+ 2 \gamma \tau_a +4m)}. \label{eq:L01_10}
\eea
Next, we move to the second diagonal $k+n=2$ which contains the entries $M(2,0), M(1,1)$ and $M(0,2)$. Setting $k=1,n=1$ and $k=0,n=2$ in  \eref{eq:rec1M} we find the equations satisfied by these quantities, which are given by,
\begin{align}
\mu M(2,0) &= m M(0,2)-\gamma M(1,1)+a_{0}L(1,0), \cr 
\mu M(1,1) &=-\gamma M(0,2)+a_{0}L(0,1).
\label{eq:M20_M11}
\end{align}
Additionally, we have $M(1,1)=0$, from \eref{eq:constr2}. 
These three equations can be solved to obtain the correlations appearing on the second diagonal, 
\bea
\la x^2 \ra &\equiv& M(2,0)= \frac{a_{0}^{2}\tau_a(\gamma\tau_a + 2 m)}{\gamma \mu (\mu\tau_a^2 + 2\gamma \tau_a +4 m)}, \cr
\la v^2 \ra &\equiv&  M(0,2) = \frac{2 a_{0}^{2} \tau_a}{\gamma(\mu\tau_a^2 +2\gamma \tau_a +4 m)},\label{eq:M02}
\label{eq:M20}
\eea 
where we have used \eref{eq:L01_10} obtained in the previous step. 
One can proceed in a recursive manner to compute the higher order correlations. In general, the $d$-th diagonal contains $d+1$ entries, which satisfy $d$ equations, obtained from \eref{eq:rec1M}-\eqref{eq:rec1L}, along with an additional constraint from \eref{eq:constr2}. These $(d+1)$ coupled equations can be solved to obtain the correlations $M(k,d-k)$ or $L(k,d-k)$. 

It is worthwhile to look at the stationary variance of the position and velocity, given by \erefs{eq:M20}. Interestingly, the position variance $\la x^2 \ra $ shows a non-monotonic behaviour as the activity $\tau_a$ increases in the regime $\tau_2 < \tau_1$. As the activity $\tau_a$ is increased, the variance first increases, attaining a maximum value at $\tau_a = 2m/(\sqrt{m \mu} - \gamma)$. The variance then decreases with $\tau_a$, eventually becoming $a_0^2/\mu^2$ asymptotically in the limit $\tau_a\to\infty$. This is in contrast with the ordinary RTP, i.e., the $m=0$ case, where the variance of the position increases monotonically with $\tau_a$. On the other hand, the velocity variance $\la v^2 \ra$ always shows a non-monotonic behaviour with $\tau_a$ with a maximum at $\tau_a = \sqrt{4 m /(\mu)}$. 

\begin{figure}[t]
\begin{centering}
\includegraphics[width=8.cm]{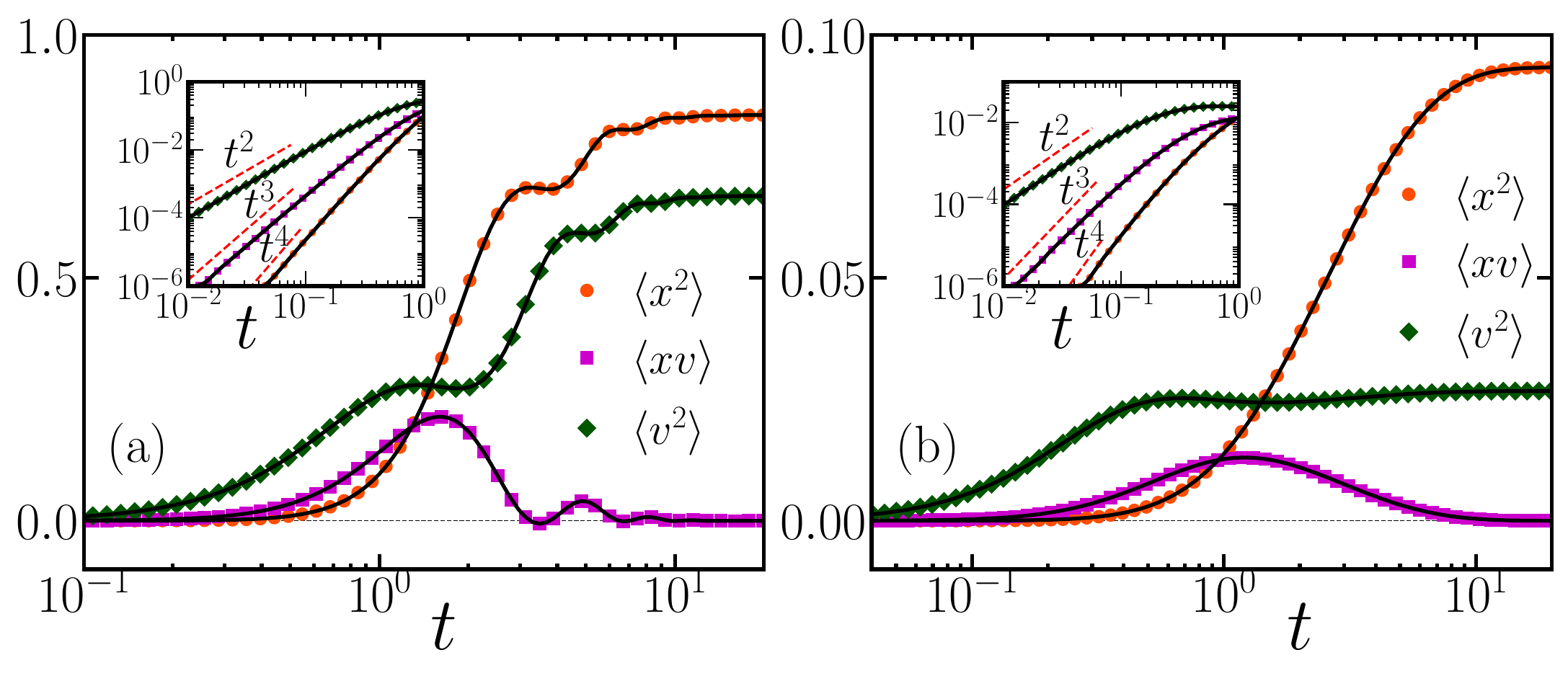}
\end{centering}
\caption{Time evolution of the moments in the (a) underdamped and (b) overdamped scenarios.  The symbols indicate the data obtained from numerical simulations while the solid lines indicate the analytical predictions [see \erefs{eq:x2t_full_od}-\eqref{eq:xvt_full_ud}].  The insets illustrate the anomalous growth at short-times [see \erefs{eq:x2_sht}-\eqref{eq:v2_sht}] using the same data plotted in log-scales. Here we have used: $\tau_a=1.0,\mu=1.0,m=1.0,\gamma=0.5$ and $a_0=1.0$ for the underdamped case, and $\tau_a=1,\mu=1,m=1.0,\gamma=5.0$ and $a_0=1.0$ for the overdamped case.}
\label{fig:x2_v2}
\end{figure}

It is also useful to look at the time evolution of the position and velocity correlations. To this end, we compute the full time-dependent correlations $\la x^2(t) \ra$, $\la v^2(t) \ra$ and $\la x(t) v(t)\ra $ from the formal solution of the Langevin equations \eqref{eq:x_formal}-\eqref{eq:v_formal}. The details of this calculation is provided in Appendix \ref{appB}, here we only mention the notable features. 
At short-times, i.e., for $t \ll \min(\tau_1, \tau_2, \tau_a)$, we have,
\begin{align}
\la x^{2}(t)\ra & = \big(\frac{a_{0}}{2 m}\big)^2 t^4 + O(t^5),\label{eq:x2_sht}\\
\la x(t)v(t)\ra & = \frac{1}{2}\big(\frac{a_0}{m}\big)^2 t^3 + O(t^4),\label{eq:xv_sht}\\
\la v^{2}(t)\ra & = \big(\frac{a_0}{m}\big)^2 t^2 + O(t^3).\label{eq:v2_sht}    
\end{align}
On the other hand, at intermediate times, the correlations show oscillatory behaviour in the underdamped case, which is absent in the overdamped case. Figure~\ref{fig:x2_v2} shows the time evolution of the second order correlations, $\la x^2 (t) \ra$, $\la x(t)v(t) \ra$ and $\la v^2 (t) \ra$ obtained from numerical simulations, using Euler discretization, along with the corresponding analytical expressions \erefs{eq:x2t_full_od}-\eqref{eq:xvt_full_ud} for the underdamped and the overdamped cases.

\section{Underdamped Case} \label{ud_case}
\noindent  
In this section, we study the marginal position and velocity distributions for the underdamped scenario $\gamma^2<4m\mu$, i.e. when $4\tau_1>\tau_2$. 
In this case, between successive tumbling events, where $\sigma$ remains constant, the particle undergoes a deterministic underdamped motion in a harmonic trap centred at $a_0 \sigma/\mu$ [see \erefs{eq:xt_ud}-\eqref{eq:vt_ud}]. Each tumbling event results in a sudden change of the position of the trap minimum, i.e., $a_0\sigma/\mu\to -a_0\sigma/\mu$, with the resulting motion now directed towards the new minimum [see Fig.~\ref{fig:traj_oud}]. It is evident from \erefs{eq:xt_ud}-\eqref{eq:vt_ud}, this underdamped run-and-tumble motion is characterized by the inertial time-scale $\tau_1=m/\gamma$ associated with the relaxation of the particle in a harmonic trap, and the active time-scale $\tau_{a}$ associated with the switching of the trap position. The interplay between these two time-scales gives rise to certain novel features for the joint stationary distribution $P_{\text{st}}(x,v)$ [see Fig.~\ref{fig:Pxv_ud_od} (d),(e)]. To understand this unique behaviour in the underdamped regime, we discuss the marginal position and velocity distributions separately.

\begin{figure}[t]
\centering
\includegraphics[width=8.0 cm]{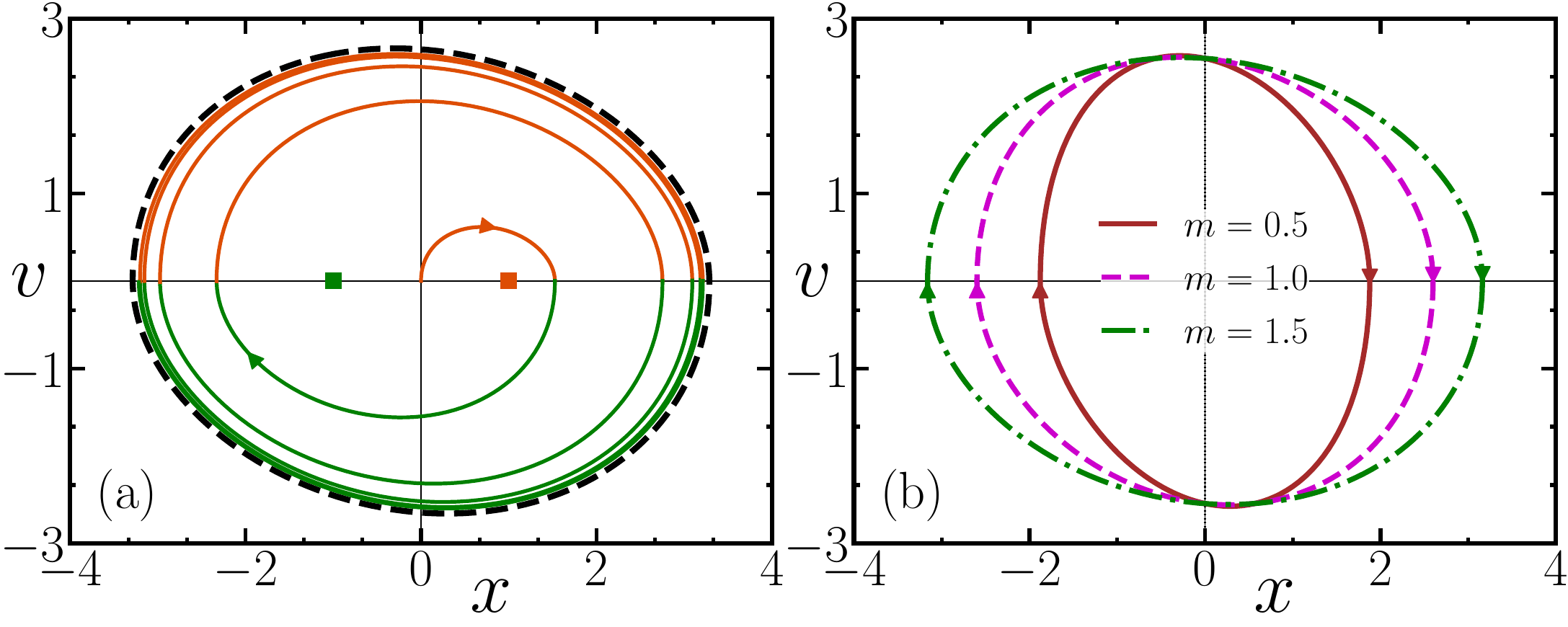}
\caption{(a) Underdamped scenario: Typical displacement maximizing trajectory of the particle shows that the particle settles into a limit-cycle (black dashed line) described by the trajectories \erefs{eq:lcyc_2}-\eqref{eq:lcyc_1}. Here we have used $\mu=1.0,a_0=1.0,\gamma=0.5$ and $m=1.5$. (b) Trajectories [see \erefs{eq:lcyc_2}-\eqref{eq:lcyc_1}] bounding the joint stationary distribution $P_{\text st}(x,v)$ for different values of mass $m$. Here we have used $\mu=1.0,a_0=1.0$ and $\gamma=0.5$.}\label{fig:bound_traj_ud}
\end{figure}

We start by computing the finite supports of the distributions which arise due to the bounded nature of the active force $\sigma(t)$. To compute these bounds we consider the trajectory which maximizes the displacement of the particle. The displacement of an underdamped particle from its initial position $x_0$ is maximum when the particle reaches the first turning point, i.e. where its velocity changes sign for the first time. Thus, for an IRTP, the displacement will be maximized if the tumbling events $\sigma \to -\sigma$ coincide with the velocity $v$ changing its sign. Hence, the bound on the position fluctuation can be obtained by considering such a trajectory [see Fig.~\ref{fig:bound_traj_ud}(a)].  

Let $x_\ell$ denote the particle position at the $\ell$-th turning point of this trajectory. By definition, the $\ell$-th turning point coincides with the $\ell$-th tumbling event, $\sigma_{\ell}\to-\sigma_{\ell-1}$. Considering the harmonic nature of the motion between successive tumbling events, it is straightforward to show that $x_{\ell}$ satisfies the recursion relation [see Appendix \ref{appC} for the details], 
\bea
x_{\ell+1}=c x_{\ell}+\frac{a_0}{\mu}(1-c)\sigma_{\ell},
\eea
where $c=-e^{-\frac{\pi}{\Lambda}}$ and $\Lambda=\sqrt{4m\mu / \gamma ^ 2-1}$ [see \eref{eq:Lamb_def}].  Without any loss of generality, we assume that the particle starts at rest from position $ |x_0| < a_0 / \mu$  with propulsion direction $\sigma_0$. Then, the solution of this recursion relation is given by,
\bea
x_{\ell} = c^{\ell}x_{0} + \frac{a_{0}\sigma_{0}}{\mu} \Big[\frac{1-c}{1+c}\Big]\big[(-1)^{\ell-1}+c^{\ell}\big],\label{eq:rec_sol}
\eea
where we have used the fact that $\sigma_\ell = (-1)^{\ell}\sigma_0$. In that case, it is straightforward to show that $|x_\ell| - |x_{\ell-1}| >0$ for all $\ell>0$, i.e., the distance of the particle from the origin increases at successive turning points. Since $c < 1$, $x_\ell$ approaches a finite value in the limit $\ell \to \infty$.
Taking this limit, $\lim_{l\to\infty}x_{\ell}$, gives the maximum distance the particle can reach, which by definition is the bound for the position distribution quoted in \eref{eq:x_b}. Note that, this bound on the position distribution of an IRTP has also been obtained earlier by considering the large deviation function~\cite{smith2022nonequilibrium}.

\begin{figure}
\centering
\includegraphics[width=8.8 cm]{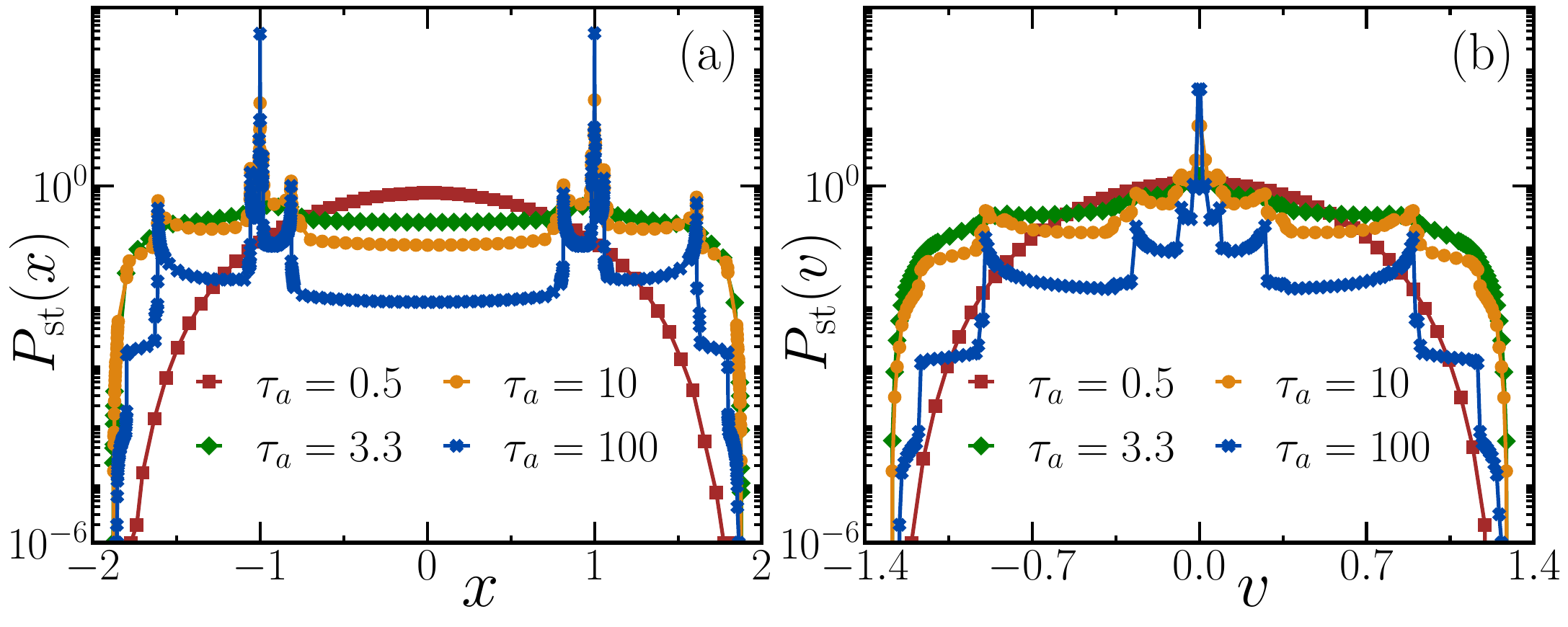}
\caption{Underdamped scenario: Plot of position distribution $P_{\text{st}}(x)$ (a) and velocity distribution $P_\text{st}(v)$, obtained from numerical simulations, for different values of activity $\tau_a$. Here we have used $\gamma=1.0,\mu=1.0$ and $m=2.0$.}
\label{fig:dist_ud_alp_xv}
\end{figure}

The bound on the velocity fluctuations can also be computed from this trajectory, following a similar argument. We obtain this bound explicitly [quoted in \eref{eq:v_b}] by computing the velocity of the particle where the acceleration changes sign in the trajectory [see Appendix \ref{appC} for the details].
Note that, in the strongly underdamped limit $m\rightarrow\infty$, the position bound $X_{\text{UD}}\sim\sqrt{m}$, i.e., position distribution becomes unbounded, whereas the bound for the velocity distribution remains finite, with $V_{\text{UD}}\to 4a_0/\gamma \pi$. 

In fact, it is easy to see that the particle motion is bounded by the trajectories which connect the points  $(\pm X_{\text{UD}}, 0)$. This allows us to compute the bound of the joint distribution by computing these trajectories parametrically. Using \erefs{eq:x_formal}-\eqref{eq:v_formal} we obtain these trajectories explicitly by setting $\sigma=\pm1$, $x_0=\mp X_{\text{UD}}$ and $v_0=0$ as the initial conditions,
\begin{align}
X_{\text{UD}}(t) &= \pm\frac{a_0}{\mu}\mp\frac{a_0}{\mu}e^{-\frac {t}{2\tau_1}}(1+\coth\frac {\pi}{2\Lambda})\cr 
&\qquad\qquad \times \big[\frac 1 \Lambda \sin \frac{\Lambda t}{2\tau_1} + \cos \frac{\Lambda t}{2\tau_1} \big],\label{eq:lcyc_2}\\
V_{\text{UD}}(t) &= \frac{2\bar{v}_0}{\Lambda}e^{-\frac{t}{2\tau_1}}\sin\frac{\Lambda t}{2\tau_1}(1+\coth\frac {\pi}{2\Lambda}),\label{eq:lcyc_1}
\end{align}
where $0\le t\le 2\tau_1\pi/\Lambda$. Figure~\ref{fig:bound_traj_ud} shows typical trajectories describing the boundary of the joint stationary distribution in the underdamped scenario for different values of particle mass. In the following, we separately discuss the behaviour of the position and the velocity distributions.

\subsection{Position Distribution}

\begin{figure}
\centering
\includegraphics[width=8.0 cm]{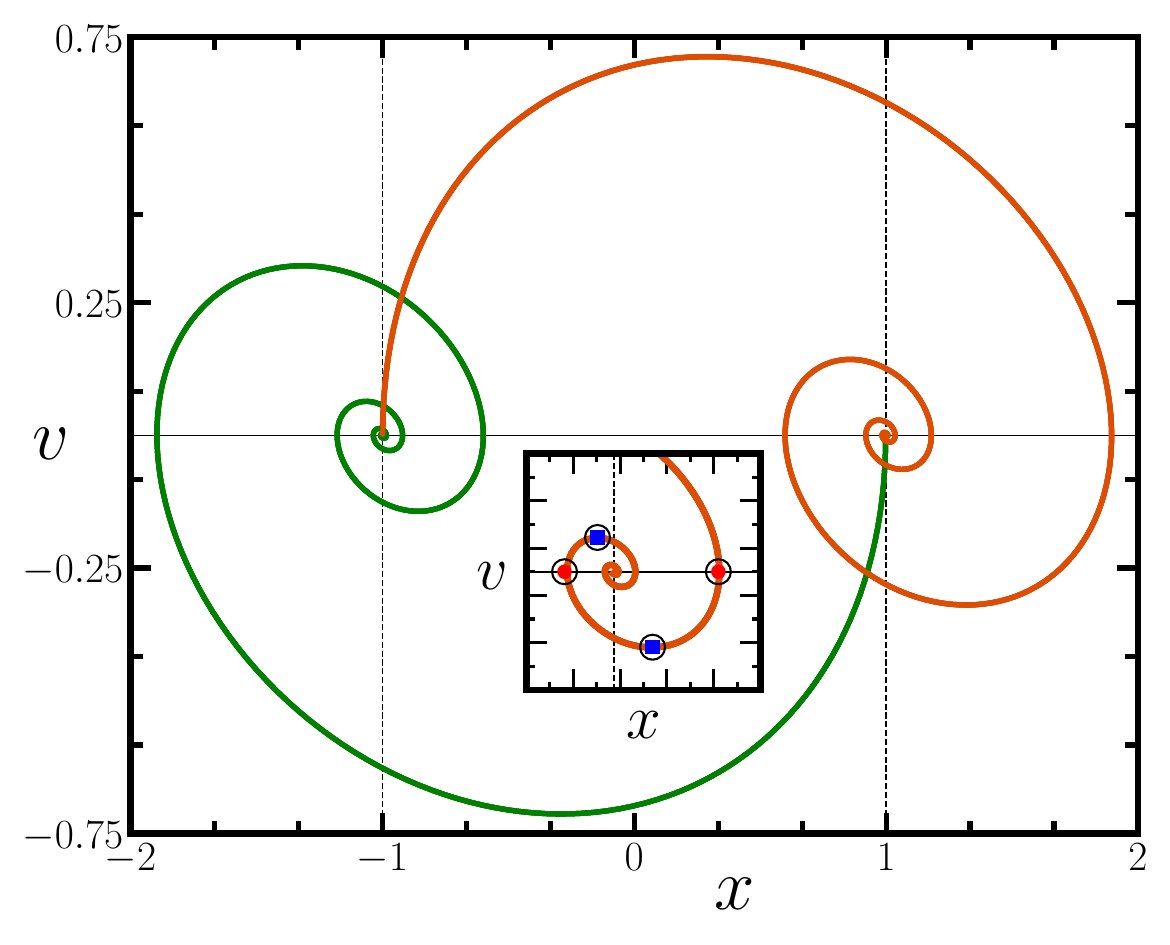}
\caption{Underdamped strongly active regime: Typical trajectory of the IRTP in the phase space. The inset shows the turning points in the trajectory marked using symbols---disks for the position turning points; and squares for the velocity turning points. Here we have used $\gamma=1.0, \mu=1.0, a_0=1.0, m=4.0$ and $\tau_a = 100$.}
\label{fig:ud_av_tp}
\end{figure}

In this section, we investigate the behaviour of the marginal position distribution of the IRTP in the underdamped case. The stationary position distribution $P_\text{st}(x)$ undergoes a shape transition with change in the activity $\tau_a$. For $ \tau_1 \ll \tau_a$, i.e., in the strongly active regime, the position distribution shows a novel shape --- it has two strong peaks at $x=\pm a_0/\mu$, accompanied by several smaller peaks and shoulder-like structures near the tails. These peaks smooth out as $\tau_a$ is decreased, eventually giving rise to a dome-shaped distribution in the passive regime $\tau_a \ll  \tau_1$. This shape transition is illustrated in Fig.~\ref{fig:dist_ud_alp_xv} (a) which shows $P_{\text{st}}(x)$, measured from numerical simulations, for different values of activity $\tau_a$.

To understand the multi-peaked shape of the position distribution in the strongly active regime, we again consider the trajectory of the IRTP. Let us recall that, for a constant $\sigma$, i.e., between two successive tumbling events, the IRTP undergoes a deterministic underdamped motion in a harmonic trap centred at $a_0 \sigma/\mu$. For $\tau_{1} \ll \tau_a$, the typical time between two successive tumbling events is much larger compared to the time required by the particle to reach the trap minimum. Thus, in the strongly active regime, the trajectory of the IRTP consists of alternate segments where an underdamped harmonic oscillator starting at rest from $x=a_0 \sigma/\mu$ relaxes to $x=-a_0 \sigma/ \mu$, and stays there until the next tumbling event [see Fig.~\ref{fig:ud_av_tp} for a typical trajectory]. Moreover, before reaching the trap minimum, the particle undergoes an oscillatory motion around it, with its velocity changing sign at the turning points. Correspondingly, the particle predominantly resides near these turning points, giving rise to the multi-peaked structure of the distribution. 

\begin{figure}
\includegraphics[width=7.7cm]{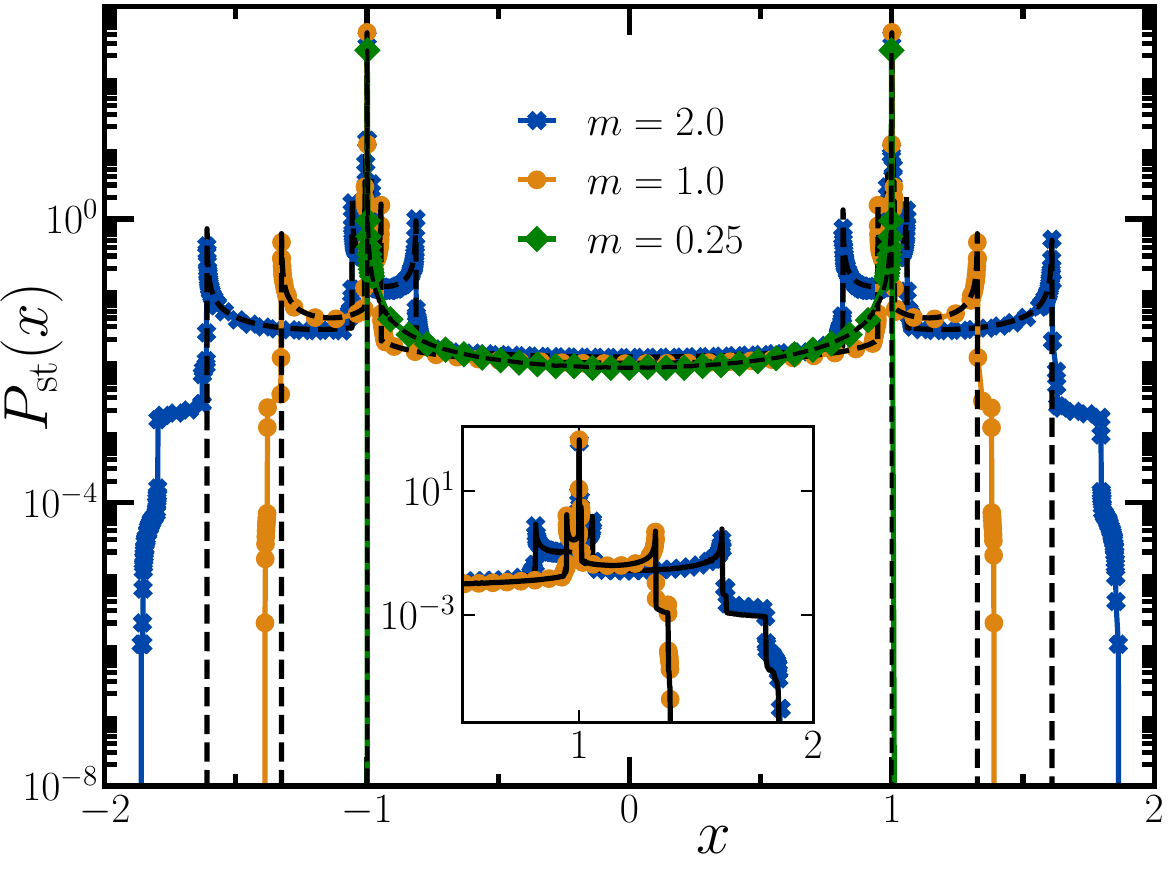}
\caption{Underdamped strongly active regime: Plot of the velocity distribution $P_\text{st}(v)$ obtained from numerical simulations of IRTP (symbols) and the approximate deterministic dynamics (dashed black lines), for different values of mass $m$. The inset compares the shoulder-like structures 
observed in IRTP (symbols) with the same obtained by incorporating a uniformly distributed waiting time distribution in the approximate deterministic dynamics (solid lines). Here we have used parameter values $\mu=1.0,\gamma=1.0,a_0=1.0$ and $\tau_a=100$.}
\label{fig:dist_low_alp_1}
\end{figure}

To quantitatively estimate this position distribution, we assume that, in the strongly active regime, the tumbling dynamics can be approximated by a deterministic one, with flips taking place at a fixed interval $\tau_a$. We measure the position distribution of this approximate deterministic dynamics using numerical simulations, which is shown in Fig.~\ref{fig:dist_low_alp_1} along with the same obtained from the actual IRTP dynamics. It is apparent that the two show an excellent agreement---the approximate dynamics correctly reproduces the multi-peaked shape of the position distribution (black dashed lines) except for the shoulder-like structures at the tails.

This approximate dynamics allows us to locate the position of the peaks, which, as mentioned already, are nothing but the turning points of the deterministic trajectory of the particle travelling between $\pm a_0/\mu$.  These can be obtained systematically by setting $v=0$ in the equation of the trajectory starting with $x_0=\pm a_0/\mu,v_0=0$ with $\sigma=\mp 1$. The details of this computation is provided in Appendix~\ref{appC}, here we only quote the main result---in the strongly active regime, the position distribution is peaked around the points, 
\bea
x^*_n=\pm \frac{a_{0}}{\mu}\Big[1-2(-1)^{n}e^{-\frac{n\pi}{\Lambda}}\Big], ~~\text{with}~~ n=1,2,3 \dots
\label{eq: peak_p}
\eea
where $\Lambda$ is defined in \eref{eq:Lamb_def}. Clearly, in the $n \to \infty$ limit, $x^*$ converges to the primary peak positions $\pm a_0/\mu$.  

Note that, this approximate dynamics ignores the effect of possible tumbling events occurring while the particle is traversing between $\pm a_0/\mu$. The effects of such events can be incorporated into the approximate dynamics by allowing the waiting times to be drawn from a uniform distribution in $[0,2\tau_a]$ which has the same mean $\tau_a$ as the original exponential distribution in the IRTP dynamics. The inset in Fig.~\ref{fig:dist_low_alp_1} shows that this inclusion of minimal stochasticity reproduces the shoulder-like structures qualitatively. 

\begin{figure}
\centering
\includegraphics[width=7.7cm]{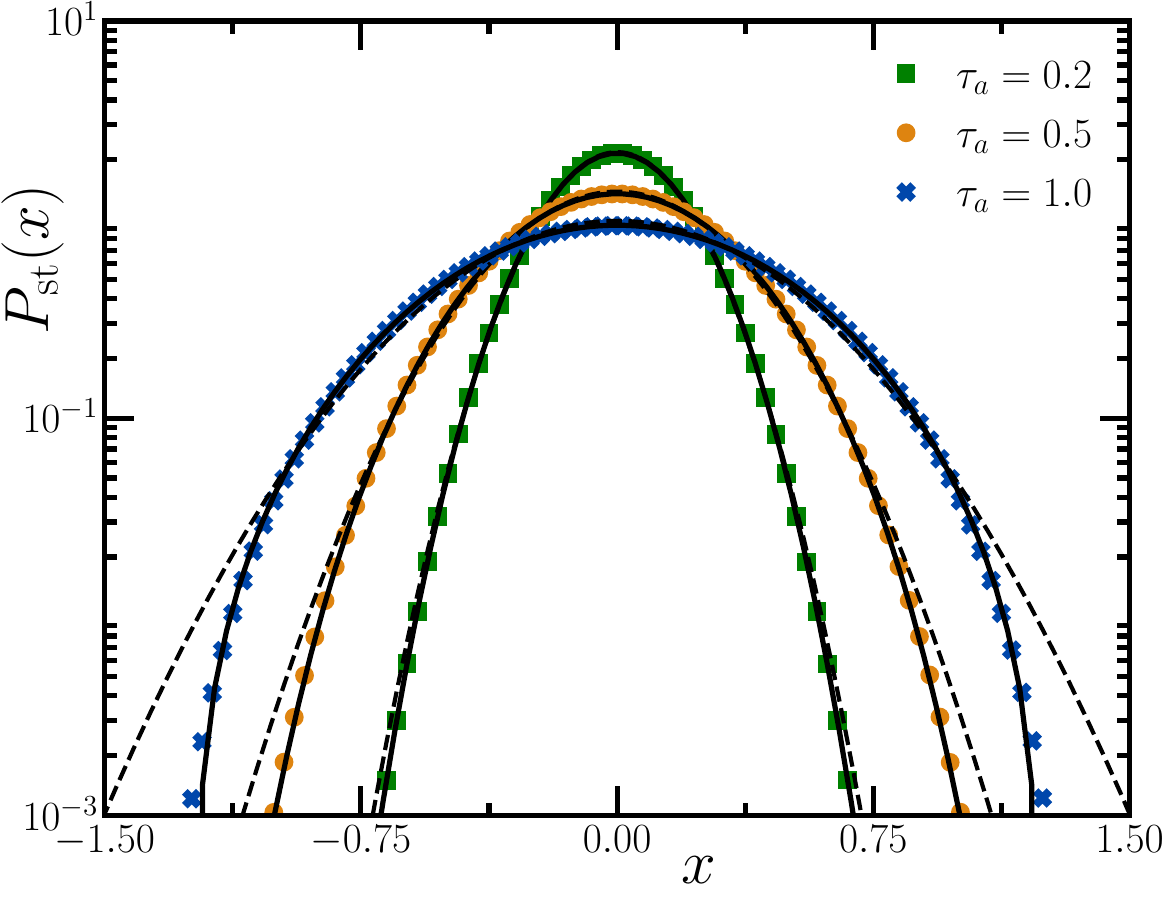}
\caption{Underdamped strongly passive regime: Plot showing the stationary position
distribution $P_{\text{st}}(x)$ of the IRTP for different values of activity $\tau_a$. The symbols are for simulation results and the solid lines stand for the corrected Gaussian given by Eq.~(\ref{eq:corr_Gauss}). Here we have used $\mu=1.0,\gamma=3.0$ and $m=10$.}
\label{fig:dist_high_alp_x}
\end{figure}

As $\tau_a$ is decreased (below $\tau_1$), the position distribution eventually becomes dome-shaped with a single peak at the origin [see Fig.~\ref{fig:dist_ud_alp_xv}]. In the strongly passive limit, i.e., $\tau_a\ll \tau_1$, the dichotomous noise $\sigma(t)$ behaves like a Gaussian white noise with strength $a_0^2/(2\gamma)$ [see \cite{santra2021active}] Correspondingly, \eref{eq:xt_ud} implies that the typical position distribution also becomes a Gaussian with width $a_0\sqrt{\tau_a/2\mu\gamma}$. For any finite $\tau_a$, corrections to the Gaussian appear at the tails. In the following, we show that the corrections to the Gaussian can be systematically obtained from the higher moments of the position. 

\begin{figure}[t]
\includegraphics[width=8cm]{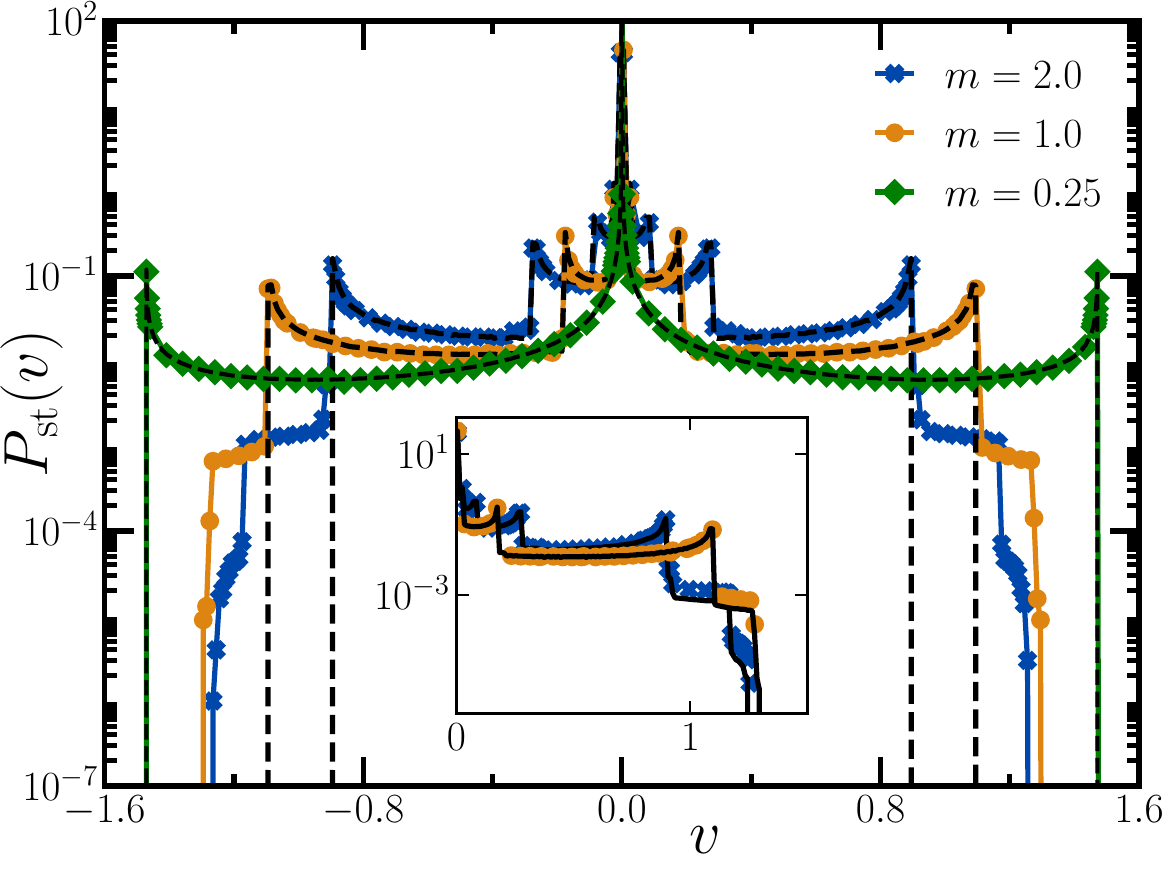}
\caption{Velocity distribution in the underdamped case in the strongly active regime: Plot of $P_\text{st}(v)$ obtained from numerical simulations of IRTP (symbols) and the approximate deterministic dynamics (dashed black lines), for different values of mass $m$. The inset compares the shoulder-like structures 
observed in IRTP (symbols) with the same obtained by incorporating a uniformly distributed waiting time distribution in the approximate deterministic dynamics (solid lines). Here we have taken $\mu=1.0,\gamma=1.0,a_0=1.0$ and $\tau_a=100$.}
\label{fig:dist_low_alp_2}
\end{figure}

To this end, we recall that the cumulant generating function (CGF) of the position distribution, which is symmetric around $x=0$, is given by,
\bea 
Q(q) = -\sum_{k=1}^{\infty}\frac{q^{2k}}{(2k)!}\langle x^{2k}\rangle_c,\label{eq:CGF}
\eea 
where $\la x^{2k} \ra_c$ denote the corresponding even cumulants. The CGF, in turn, is related to the moment-generating function,  $\tilde{P}(q)=\la e^{iqx}\ra$ by,
\bea
Q(q) = \log \tilde P(q) \label{eq:MGF}.
\eea
If all the stationary cumulants of the distribution are known, then the exact stationary distribution can, in principle, be computed using \erefs{eq:CGF}-\eqref{eq:MGF}, and taking the inverse Fourier transform of $\tilde P(q)$. 
Clearly, if one considers only $k=1$ term in \eref{eq:CGF},  up to the variance $\la x^2 \ra$, this leads to a Gaussian distribution. Corrections to this Gaussian distribution can be systematically computed by truncating \eref{eq:CGF} by keeping upto $k=N$ terms. 

To this end, we compute the first $N$ non-zero cumulants of the position distribution from the corresponding moments, which, in turn, are computed using the recursive procedure illustrated in Sec. \ref{sec:moments}. Using this truncated CGF, we then calculate $\tilde P(q)$ from \eref{eq:MGF}. Numerically inverting the Fourier transform gives us the position distribution. 
\bea
P_{\text{st}}(x)=\intop_{-\infty}^{\infty}\frac{dq}{2\pi}~\tilde{P}(q)\,\cos{qx}.
\label{eq:corr_Gauss}
\eea
Figure \ref{fig:dist_high_alp_x} shows the distribution obtained keeping $N=5$ terms, which shows an excellent agreement with the data obtained from numerical simulations. Note that, the $P_{\text{st}}(x)$ shows significant deviation from the Gaussian as the activity $\tau_a$ is increased. 

It should be noted that, this method works well only in the passive regime, where the  contributions from the higher order cumulants decrease progressively. 
In the active regime, where the distribution has a multiple peaks away from the origin, significant contributions arise from the higher order cumulants, and this method is not expected to work. 

\subsection{Velocity Distribution}\label{Pv_ud}

\begin{figure}[t]
\centering
\includegraphics[width=7.7cm]{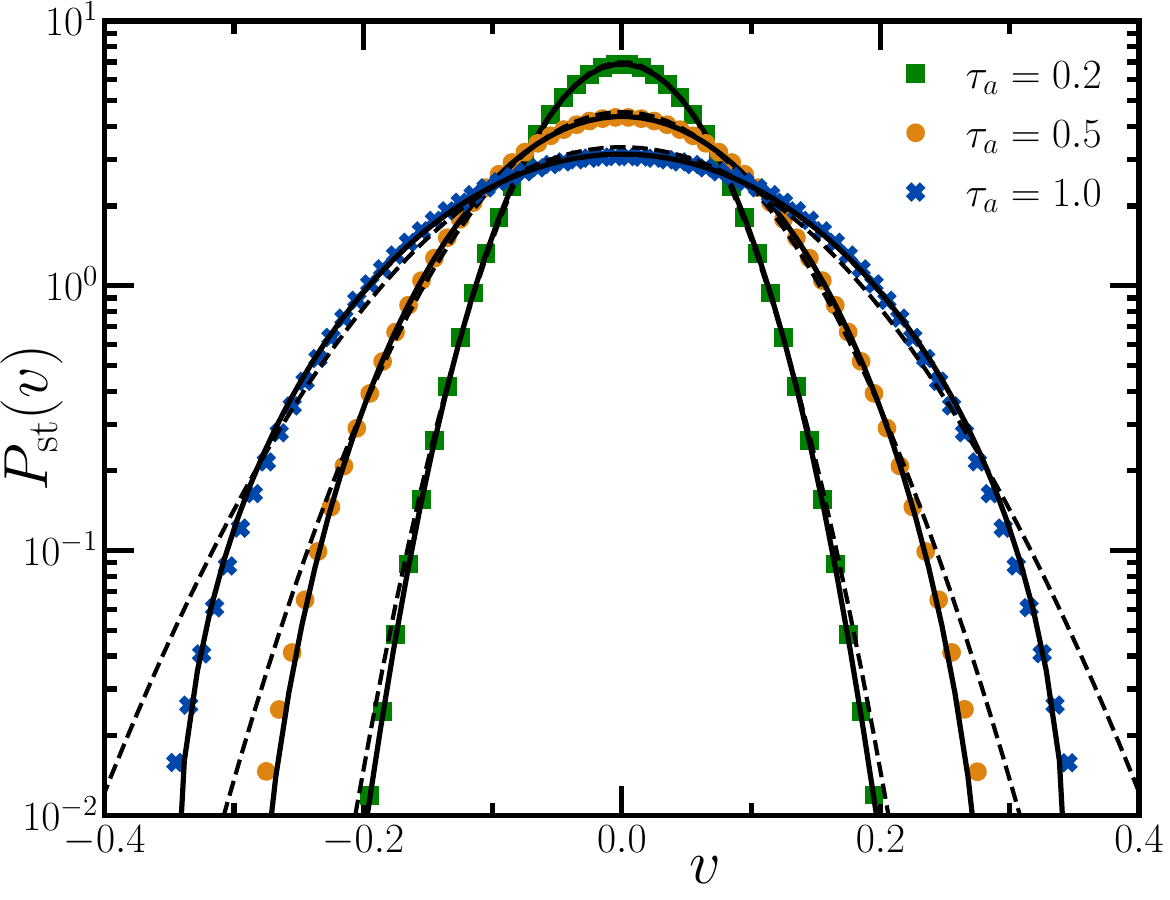}
\caption{Underdamped strongly passive regime: Plot showing the stationary velocity
distribution $P_{\text{st}}(v)$ of the IRTP for different values of activity $\tau_a$. The symbols are for simulation results and the solid lines stand for the corrected Gaussian given by \eref{eq:corr_Gauss_}. Here we have used $\mu=1.0,\gamma=3.0$ and $m=10$.}
\label{fig:dist_high_alp_v}
\end{figure}

In this section, we investigate the nature of the velocity fluctuations in the underdamped scenario. The stationary velocity distribution also shows a shape transition similar to the position distribution--- for $\tau_1 \ll \tau_a$, i.e., in  the strongly active regime, the distribution has a primary peak at $v=0$, with numerous smaller peaks, and shoulder-like structures near the tail. Decreasing activity $\tau_a$, is marked by the smoothing of the peaks, till a dome-shaped distribution emerges in the passive regime  $\tau_a \ll \tau_1$. This shape transition is illustrated in Figure~\ref{fig:dist_ud_alp_xv}(b) which shows $P_\text{st}(v)$, measured from numerical simulations, for different values of activity $\tau_a$. 

The novel multi-peaked structure of the velocity distribution in the strongly active regime can also be understood using the trajectory-based picture, similar to the position distribution. As already discussed, in this limiting scenario, the typical time between two successive tumbling events is much larger than the time required to travel between the points $(\pm a_0/\mu,0)$. From Fig.~\ref{fig:ud_av_tp}, it is clear that the velocity also undergoes oscillations before the particle eventually comes to rest at the trap minimum $a_0\sigma/\mu$. Intuitively, one expects the velocity distribution to peak around the points where the acceleration is minimum, i.e., where $\dot{v}(t)=0$.

The deterministic dynamics approximation discussed in the context of position distribution can also be used to estimate the velocity distribution quantitatively. We measure the velocity distribution by numerically simulating this approximate dynamics, which is shown in Fig.~\ref{fig:dist_low_alp_2}, along with the same obtained from the actual IRTP dynamics. Clearly, the deterministic dynamics  reproduces the multi-peaked shape of the velocity distribution fairly accurately (black dashed lines), except for the shoulder-like structures at the tail.

As before, we use this deterministic dynamics to locate the peaks in the velocity distribution. These can be obtained by setting $\dot{v}(t)=0$ in the trajectory starting with $x_0=\pm a_0/\mu,v_0=0$ and for $\sigma=\mp 1$ [see appendix \ref{appC}], and are given by,
\begin{align}
v^{\star}_n &= \pm\frac{2a_0}{\sqrt{m\mu}}\exp\Big[-\frac{1}{\Lambda} \big(n\pi + \tan^{-1}\Lambda\big)\Big] \cr
&\hspace{3cm}~~\text{with}~~ n=0,1,2 \dots \label{eq:ud_activ_vd_peaks}    
\end{align}
Moreover, the shoulder-like structures near the tails in $P_\text{st}(v)$ can also be qualitatively reproduced by allowing tumbling events to occur while the particle is traversing, by drawing the waiting times from the uniform distribution in $[0,2\tau_a]$, similar to the position distribution. This is illustrated in the  inset of Fig.~\ref{fig:dist_low_alp_2}.

\begin{figure}[t]
\centering
\includegraphics[width=8.8 cm]{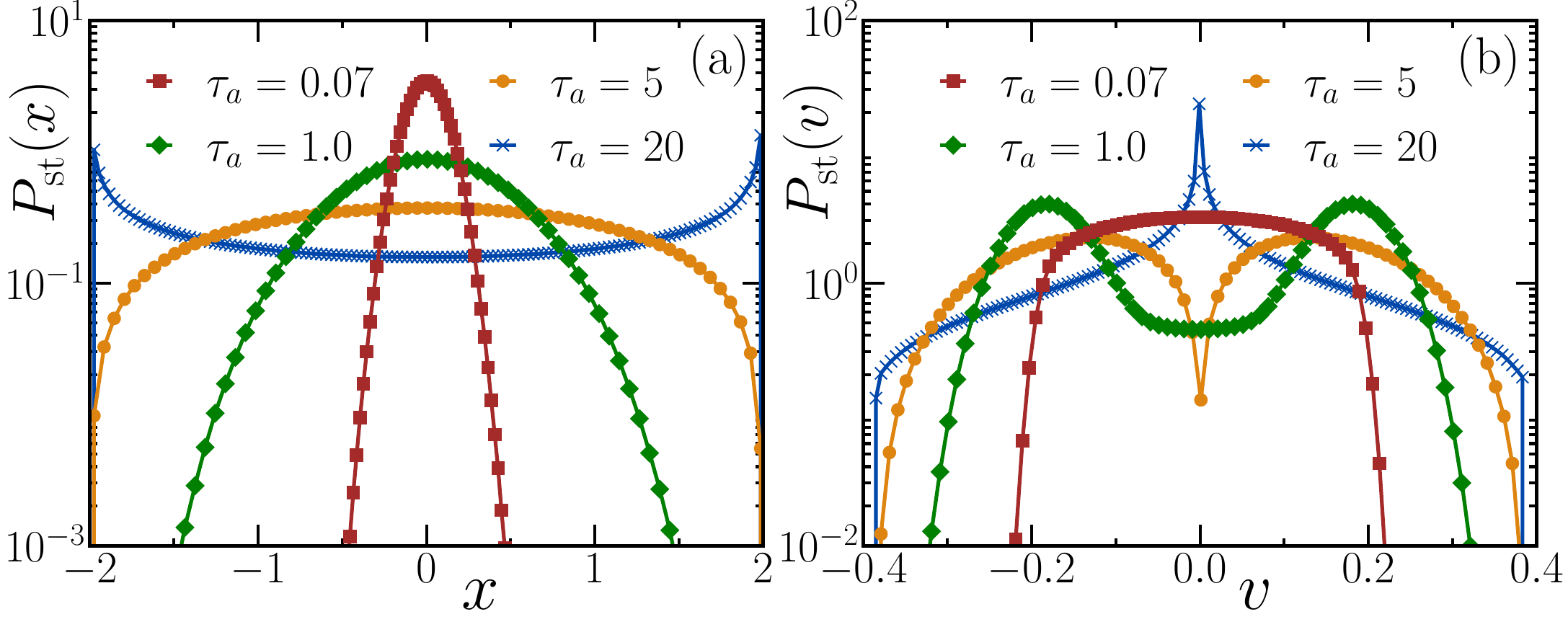}
\caption{Overdamped scenario: Plot of the position distribution $P_{\text{st}}(x)$ (a) and the velocity distribution $P_\text{st}(v)$, obtained from numerical simulations, for different values of activity $\tau_a$. Here we have used $\mu=0.5, m=0.5, \gamma=5$ and $a_0=1.0$.}
\label{fig:dist_od_alp_xv}
\end{figure}

We now turn to the strongly passive regime $\tau_a \ll \tau_1$. In this regime, the typical velocity distribution approaches a Gaussian. Similar to the position distribution, the corrections to the Gaussian, for any non-zero $\tau_a$, can be systematically obtained by considering the contributions from higher order cumulants of the velocity, in the CGF,
\bea
Q_{v}(q)=-\sum_{k=1}^{\infty}\frac{q^{2k}}{(2k)!}\la v^{2k}\ra_{c}.
\eea
Consequently, the velocity distribution can be computed by taking an inverse Fourier transform of the velocity moment-generating function, which is related to the CGF by,
\bea
Q_{v}(q)=\log\tilde{P}_{v}(q).
\eea
To this end, we construct a truncated CGF by computing the first $N$ non-zero cumulants of velocity from the corresponding moments. These moments, in turn, are obtained using the recursive procedure illustrated in Sec.~\ref{sec:moments}. Numerically inverting the Fourier transform gives us the velocity distribution,
\bea
P_{\text{st}}(v)=\intop_{-\infty}^{\infty}\frac{dq}{2\pi}\tilde{P}_{v}(q)\cos qv.\label{eq:corr_Gauss_}
\eea
Figure~\ref{fig:dist_high_alp_v} compares the velocity distribution obtained using numerical simulations, with the analytical prediction \eqref{eq:corr_Gauss_} with $N=5$, which shows an excellent agreement. 

\section{Overdamped Case}\label{sec:OD}

The overdamped scenario emerges for $\gamma^2\ge4m\mu$, i.e., when $4\tau_1\le\tau_2$. In the fully overdamped scenario, which arises in the $\tau_1 \to 0$ limit, the IRTP dynamics reduces to that of an ordinary RTP with self-propulsion speed $\bar{v}_0=a_0/\gamma$ in a trap of strength $\bar{\mu}=\mu/\gamma$. The stationary position distribution, in this limit, is bounded in the interval $[-a_0/\mu, a_0/\mu]$, and is given by~\cite{dhar2019run},
\bea
P_{\text{st}}(x)=\frac{2^{1 - 2\gamma /(\tau_a \mu)}}{B(\frac {\gamma} {\tau_a\mu}, \frac {\gamma} {\tau_a\mu})}\frac{\mu}{a_0}\Big[1-\Big(\frac{\mu x}{a_0}\Big)^2\Big]^{\frac{\gamma}{\tau_a\mu}-1},\label{eq:Px_OOD}
\eea
where $B(u, u)$ is the beta function. This distribution undergoes a shape transition at the critical activity $\tau_a^{\star}=\tau_2=\gamma/\mu$. In the active regime $\tau_2<\tau_a$, the distribution has a U-shape with algebraic divergences occurring at the two ends $x=\pm a_0/\mu$. On the other hand, in the passive regime $\tau_a<\tau_2$, a dome-shaped distribution emerges with $P_\text{st}(x)$ vanishing at the two ends. 

\begin{figure}[t]
\centering
\includegraphics[width=7.0 cm]{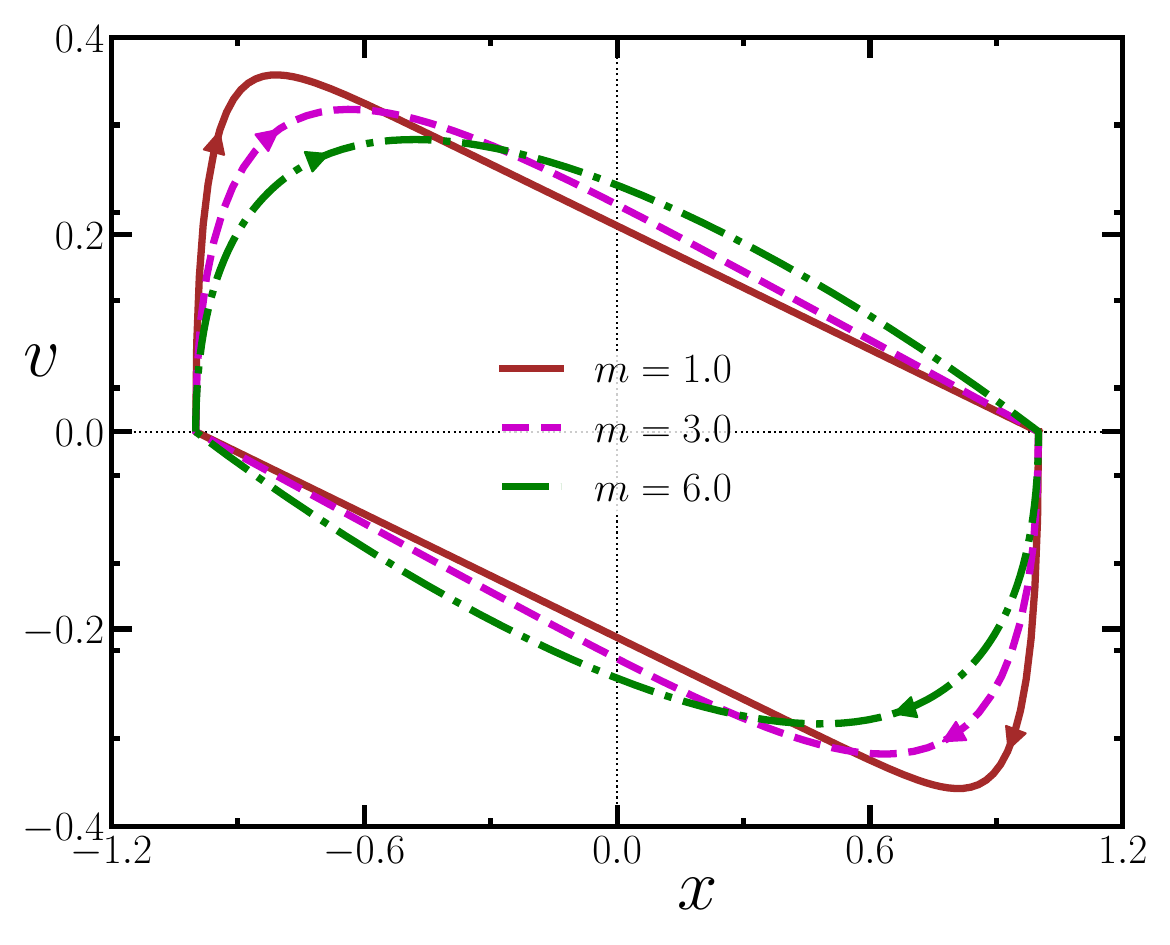}
\caption{Overdamped scenario: Typical phase trajectories bounding the joint stationary distribution $P_{\text st}(x,v)$ for different values of mass $m$ using analytic predictions from \erefs{eq:lcyc_3}-\eqref{eq:lcyc_4}. Here we have used $\mu=1.0, \gamma=5.0$ and $a_0=1.0$.}
\label{fig:bound_od}
\end{figure}

In this section, we investigate how the presence of a finite mass $m$ affects the position and velocity distributions in the overdamped regime of an IRTP. We start by investigating how the finite support of the stationary distribution changes in the overdamped scenario. A typical trajectory of the overdamped IRTP is shown in Fig.~\ref{fig:traj_oud}. Clearly, in this case, the maximum displacement of the particle corresponds to the trajectory with no tumbling events which eventually takes the particle to the trap minimum $a_0 \sigma/\mu$, without allowing it to cross it. This implies that, in the stationary state, the position of the particle is bounded in the region $-a_0/\mu<x<a_0/\mu$. A more rigorous proof of this statement is provided in Appendix \ref{appC}. This bound can also be obtained taking the limit $m\to\gamma^2/(4\mu)$, in \erefs{eq:x_b}, which gives \eref{eq:x_b_}. It is noteworthy that the position bound in this case does not depend on the mass of the IRTP.

\begin{figure}
    \includegraphics[width = 8cm]{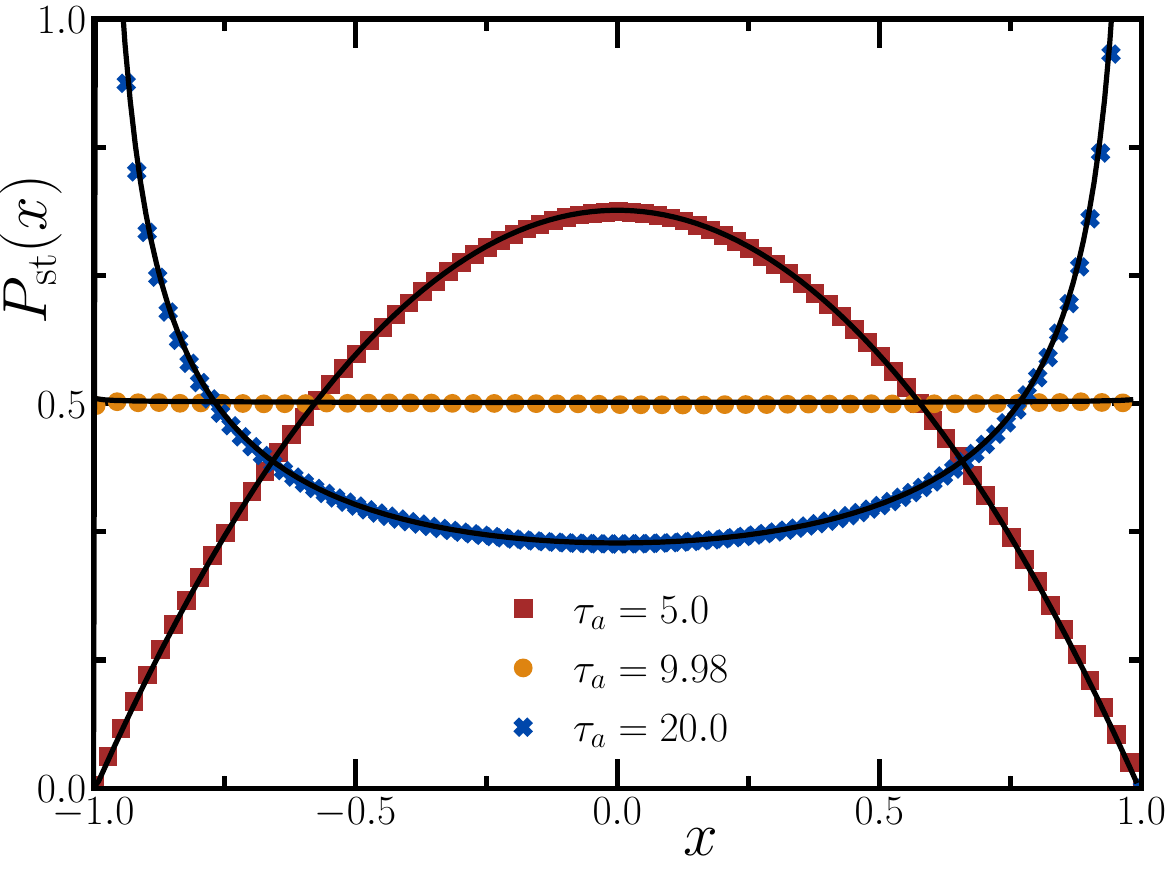}
    \caption{Overdamped scenario: Plot of the position distribution $P_{\text{st}}(x)$ obtained from numerical simulations of the IRTP (symbols) along with the approximate distribution given by Eq. (\ref{eq:Px_eff}) (solid line) for different values of activity $\tau_a$. Here we have taken $m=0.2, \mu=1.0, a_0=1.0$ and $\gamma=10$.}
    \label{fig:add_2RTP_Px}
\end{figure}

The bound in the phase-space in the overdamped case can be obtained following an argument similar to the one used in the underdamped case. 
The particle motion must be bounded by the trajectories connecting the phase points $(\pm X_\text{OD},0)$ and the velocity bound is given by the maximum velocity attained by the particle in these trajectories. By setting $\sigma=\pm 1,x_0=\mp X_\text{OD}$ and $v_0=0$ in \eref{eq:v_formal}, the bound on the velocity fluctuations is explicitly obtained  [see appendix \ref{appC} for details] 
which is quoted in \eref{eq:v_b_}. Interestingly, unlike the position, the velocity bound explicitly depends on the mass of the particle.  The boundary for the joint distribution on the $x-v$ plane is given by setting $\sigma=\pm1,x_0=\mp X_{OD}$ and $v_0=0$ simultaneously in \erefs{eq:x_formal}-\eqref{eq:v_formal},
\begin{align}
X_{\text{OD}}(t) &= \pm \frac{\bar{v}_0}{\tau_2}\Bigg[1-2 e^{-{t}/{(2\tau_1)}} \left(\cosh \frac{\lambda t}{2\tau_1}+\frac{1}{\lambda }\sinh \frac{\lambda t}{2\tau_1}\right) \Bigg],\label{eq:lcyc_3}\\
V_{\text{OD}}(t) &= \pm \frac{4\bar{v}_0}{\lambda }e^{-{t}/(2\tau_1)}\sinh\frac{\lambda t}{2\tau_1}\label{eq:lcyc_4},
\end{align}
where $ 0 \le t < \infty.$  Fig.~\ref{fig:bound_od} shows typical trajectories describing the boundary of the joint stationary distribution in the overdamped scenario for different values of particle mass.

We now turn to the marginal distributions in the overdamped case. Figure~\ref{fig:dist_od_alp_xv}(a) shows $P_{\text{st}}(x)$ measured from numerical simulations, for different values of activity $\tau_a$. Clearly, the position distribution of the IRTP shows qualitatively similar behaviour to the ordinary RTP---it undergoes a transition from a U-shape to a dome shape as the activity is decreased. However, the presence of a finite mass, i.e., non-zero $\tau_1$ changes the distribution quantitatively, including the critical value of activity, which now depends on the mass of the particle. Figure~\ref{fig:dist_od_alp_xv}(b) illustrates the velocity distribution $P_\text{st}(v)$ of the IRTP measured from numerical simulations for different activity $\tau_a$. As evident, velocity distribution, shows a very different behaviour than in the underdamped case [see Fig.~\ref{fig:dist_ud_alp_xv}(b)]. In particular, it undergoes two distinct shape transitions with decreasing activity, first from a unimodal distribution peaked at $v=0$ to a bimodal distribution with minimum at $v=0$, and then from the bimodal distribution to a Gaussian-like distribution. To further understand these behaviours in the overdamped regime, we consider the case of the strongly overdamped scenario $\tau_1\ll\tau_2$.

From \erefs{eq:xt_od}-\eqref{eq:vt_od}, it is apparent that the particle motion in the overdamped scenario is characterized by two inertial time-scales $\tau_1/(1+\lambda )$ and $\tau_1/(1-\lambda )$, where $\lambda $ is defined in \eref{eq:w_def}, along with the active time-scale $\tau_a$. In the strongly overdamped regime, i.e., for $ \gamma^2 \gg 4 m\mu$ the inertial time-scales reduce to $\tau_1=m/\gamma$ and $\tau_2=\gamma/\mu$ which are well separated. Consequently, three regimes of activity emerge, namely, strongly active $\tau_a\gg\tau_2\gg\tau_1$, moderately active regime $\tau_2 \gg \tau_a \gg \tau_1$ and the passive regime $\tau_2\gg\tau_1\gg\tau_a$. In the following, we separately explore the behaviour of the position and velocity distributions in these regimes.

\begin{figure}
\centering
\includegraphics[width = 8cm]{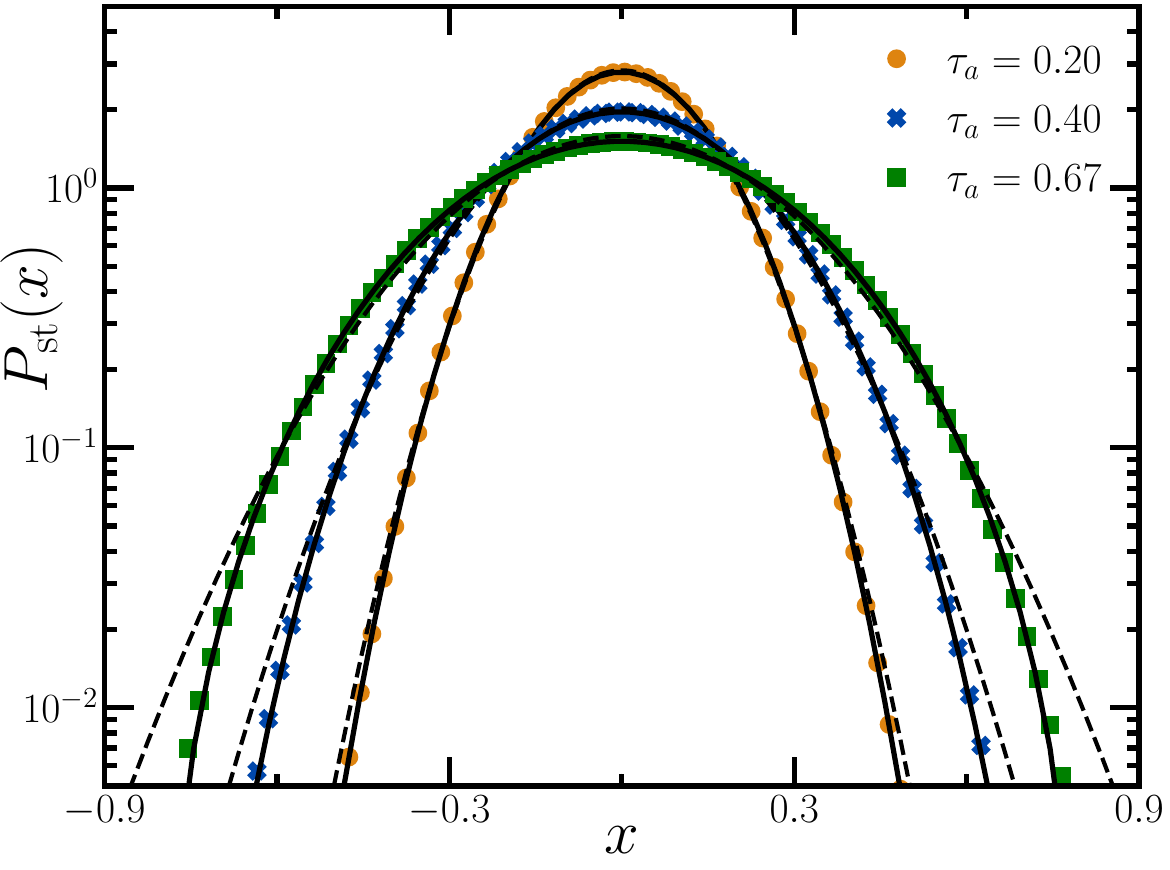}
\caption{Overdamped strongly passive regime: Plot showing the stationary position
distribution $P_{\text{st}}(x)$ for different values of activity $\tau_a$. The symbols are for simulation results and the solid lines stand for the corrected Gaussian given by Eq. (\ref{eq:corr_Gauss}). Here we have used $m=0.5,\mu=1$ and $\gamma=5$.}
\label{fig:dist_high_alp_x_od}
\end{figure}

\subsection{Position Distribution}\label{sec:OD_Pos}

\begin{figure}[t]
\centering
\includegraphics[width = 8cm]{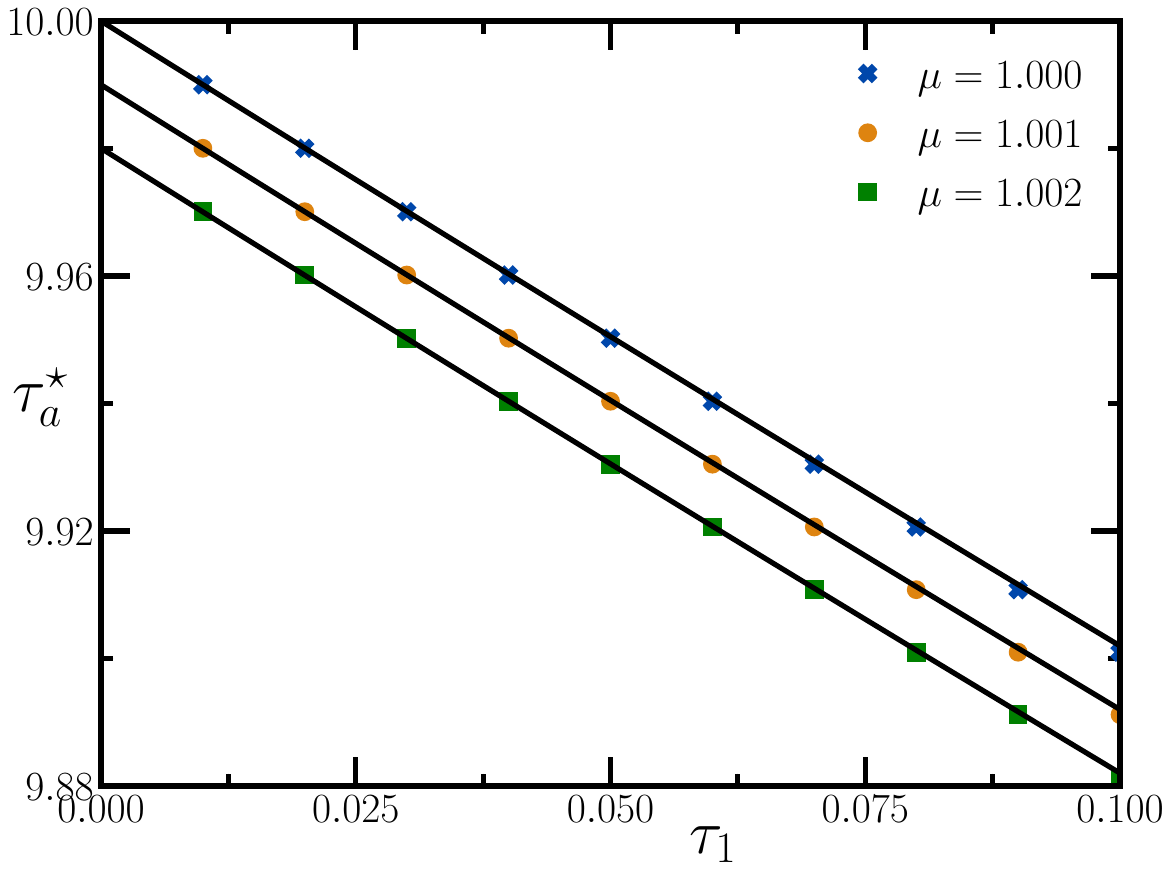}
\caption{Plot showing the dependence of the shape transition point $\tau_{a}^{\star}$ of the stationary position distribution on the viscous time-scale $\tau_1$ for different values of trap strength $\mu$ and for $a_0=1.0, \gamma=10$.}\label{fig:od_tran}
\end{figure}

In this section, we analytically characterize the position distribution in the strongly overdamped regime $\tau_1 \ll \tau_2$. We start from \eref{eq:xt_od}, the formal solution for $x(t)$ in the overdamped case. It is straightforward to see that \eref{eq:xt_od} can be recast as,
\bea
x(t) = x_{1}(t)-x_{2}(t),
\eea
where, $\{x_{1}, x_{2}\}$ are two ordinary RTPs, propelled by the same active noise $\sigma(t)$ with strength $\bar{v}_0/\lambda $, but moving in two harmonic traps of strengths $\mu_{1}=(1-\lambda )/2\tau_1$ and $\mu_{2}=(1+\lambda )/2\tau_1$, respectively.
Thus, $x_1(t)$ and $x_2(t)$ evolve according to the Langevin equations,
\begin{align}
\dot x_1(t) &= - \mu_1 x_1(t) + \bar{a}_{0} \sigma (t), \cr 
\dot x_2(t) &= - \mu_2 x_2(t) + \bar{a}_{0} \sigma (t). \label{eq:OD_recast}
\end{align}
It is convenient to introduce the variable $z(t)\equiv x_{1}(t)+x_{2}(t)$, and write the Langevin equations for $x(t)$ and $z(t)$ using \eref{eq:OD_recast},
\begin{align}
    \dot{x}(t)=&-k_+ x(t) + k_- z(t),\label{eq:xz_coup_1}\\ 
    \dot{z}(t)=&-k_+ z(t) + k_- x(t) + 2\bar{a}_0\sigma(t), \label{eq:xz_coup_2}
\end{align}
where, 
\bea
k_\pm=\frac 12 (\mu_{2}\pm\mu_{1}). \label{eq:kpm_def} 
\eea
Using the above equations, we compute the distribution of $x$ in the strongly overdamped limit ($\tau_2\ll \tau_1$) perturbatively, to the leading order in $\tau_1$. The details of this computation is provided in the Appendix~\ref{App:sto_pxv}, here we sketch the main steps.
From \eref{eq:xz_coup_1}, we can express $x$ in terms of $z$ and $\dot z$,
\begin{align}
x(t)=z(t)-2\Big(\frac{z(t)}{\tau_2}+\dot{z}\Big)\tau_1+O(\tau_1^2).\label{eq:x_m_expd}
\end{align}
Moreover, we find that, to this order, $z$ satisfies a Langevin equation,
\bea
\dot{z}(t)\approx-\kappa z(t)+\nu\sigma(t),\label{eq:zeff}
\eea
where, 
\bea
\kappa\approx\frac{1}{\tau_2}\Big(1+\frac{2\tau_1}{\tau_2}\Big),~~\text{and}~~\nu=\bar{v}_0\tau_2\kappa.\label{eq:kappa_nu_def}
\eea
Clearly, \eref{eq:zeff} describes the motion of an ordinary RTP in a harmonic trap of strength $\kappa$ and self-propulsion speed $\nu$. The corresponding joint probability distribution $P_\sigma(z)$ is known exactly \cite{dhar2019run} which can be used to obtain the distribution of $x$ [see Appendix \ref{App:sto_pxv} for the details],

\begin{align}
    P_{\text{st}}(x)&\approx\frac{(\kappa/(2\nu))^{2\alpha}}{B(\alpha, \alpha)} \Bigg[\frac{\big(X_{-}-x\big)^{\alpha-1}}{\big(X_{+}+x\big)^{-\alpha}}
    \Theta\Big(X_{+}+x\Big)\Theta\Big(X_{-}-x\Big)\cr
    &+\frac{\big(X_{-}+x\big)^{\alpha-1}}{\big(X_{+}-x\big)^{-\alpha}}\Theta\Big(X_{+}-x\Big)\Theta\Big(X_{-}+x\Big)\Bigg],\label{eq:Px_eff}
\end{align}
where $\alpha = 1/(\tau_a \kappa)$ and $b=2\bar{v}_0\tau_1(1-2\tau_1/\tau_2)$ and $\Theta(u)$ indicates the Heaviside Theta function, and we have defined,
\begin{align}
 X_\pm = X_\text{OD}\left[1 \pm \frac{2\tau_1}{\tau_2} \right].  
\end{align}
\begin{figure}[t]
    \includegraphics[width = 8cm]{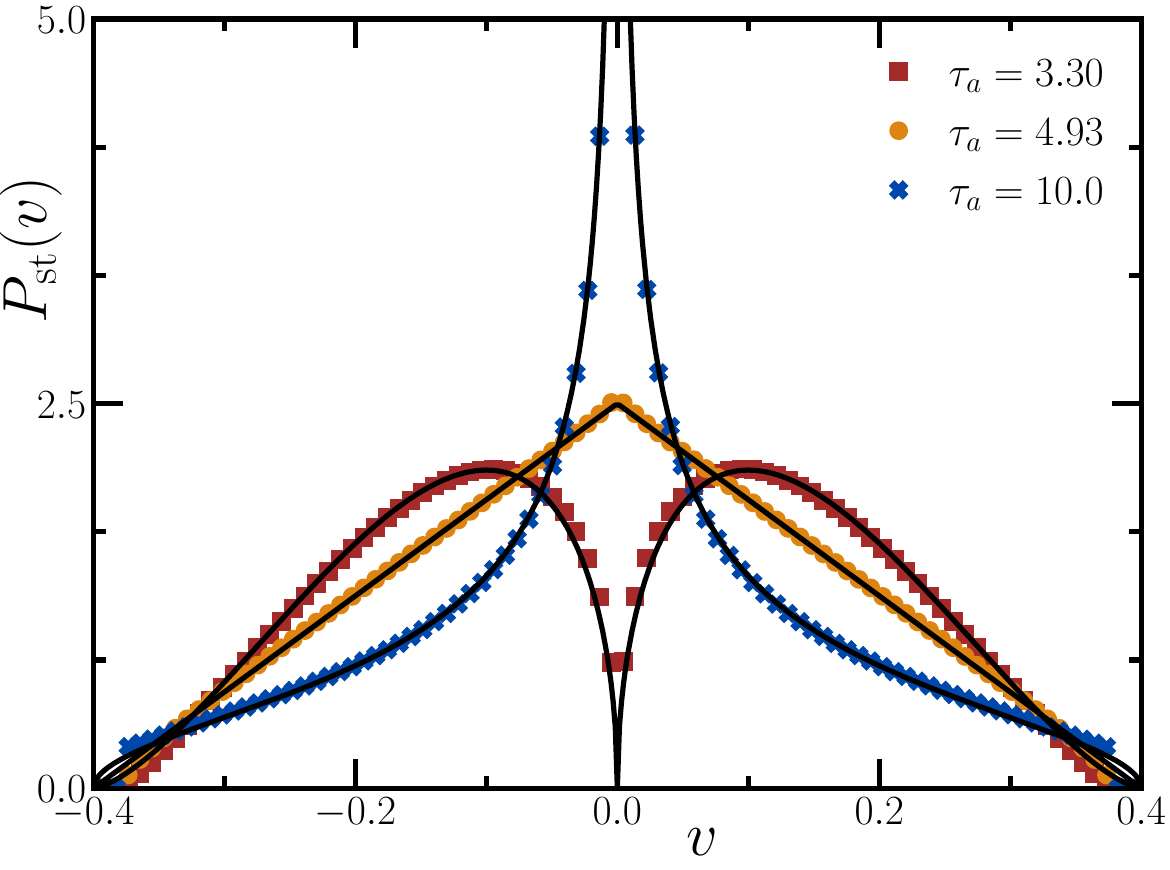}
    \caption{Overdamped active regime: Plot of the velocity distribution $P_{\text{st}}(v)$ obtained from numerical simulation of the IRTP (symbols) along with the prediction obtained from the effective model [see Eq. (\ref{eq:Pv_eff})] (solid line) for different values of activity $\tau_a$. Here we have taken $m=0.5, \mu=1.0, a_0=1.0$ and $\gamma=5$.}
    \label{fig:add_2RTP_Pv}
\end{figure}
The above equation implies that, to this order, the position distribution is bounded in the region $|x|<X_{+}$. However, as we have shown before, the exact bound for the position distribution is given by $X_\text{OD}=(X_{+}+X_{-})/2$. Hence, the approximation \erefs{eq:x_m_expd}-\eqref{eq:zeff} leads to $O(\tau_1)$ deviation from the exact bound of the position distribution. Nevertheless, we expect \eref{eq:Px_eff} to accurately describe the position distribution of the IRTP in this regime for $x$ away from $X_\text{OD}$. In fact, \eref{eq:Px_eff} obviously predicts a shape transition for the distribution at the critical activity, 
\bea
\tau_a^{*}=\kappa^{-1}=\tau_2\Big[1-2\frac{\tau_1}{\tau_2}+O(\tau_1^2)\Big]\label{eq:taus_m}.
\eea
Hence, in the active regime $\tau_a>\tau_a^{*}$, the distribution diverges at $x=\pm X_{-}$, while in the passive regime $\tau_a<\tau_a^{*}$ the distribution vanishes at $x=\pm X_{-}$. Clearly, the presence of inertia decreases the value of the critical activity.

The approximate distribution given by \eref{eq:Px_eff} is compared with the data from numerical simulations in Fig.~\ref{fig:add_2RTP_Px}, for a small value of $\tau_1$ which shows an excellent agreement. The position distribution indeed shows the shape transition predicted by \eref{eq:Px_eff}, from a U-shaped curve for $\tau_a>\tau_a^{*}$ to a inverted U-shaped curve for $\tau_a<\tau_a^{*}$. We also numerically estimate $\tau_a^{*}$ for different values of $m$, i.e., $\tau_1$ and compare it with the prediction \eref{eq:taus_m} [see Fig.~\ref{fig:od_tran}]. 

\subsection{Velocity Distribution}
In this section, we analytically characterize the position distribution in the strongly overdamped regime $\tau_1 \ll \tau_2$. It turns out that, unlike the position distribution, the velocity distribution shows qualitatively different behaviour in the strongly active, moderately active and passive regimes. 
\begin{figure}[t]
    \centering
    \includegraphics[width=8cm]{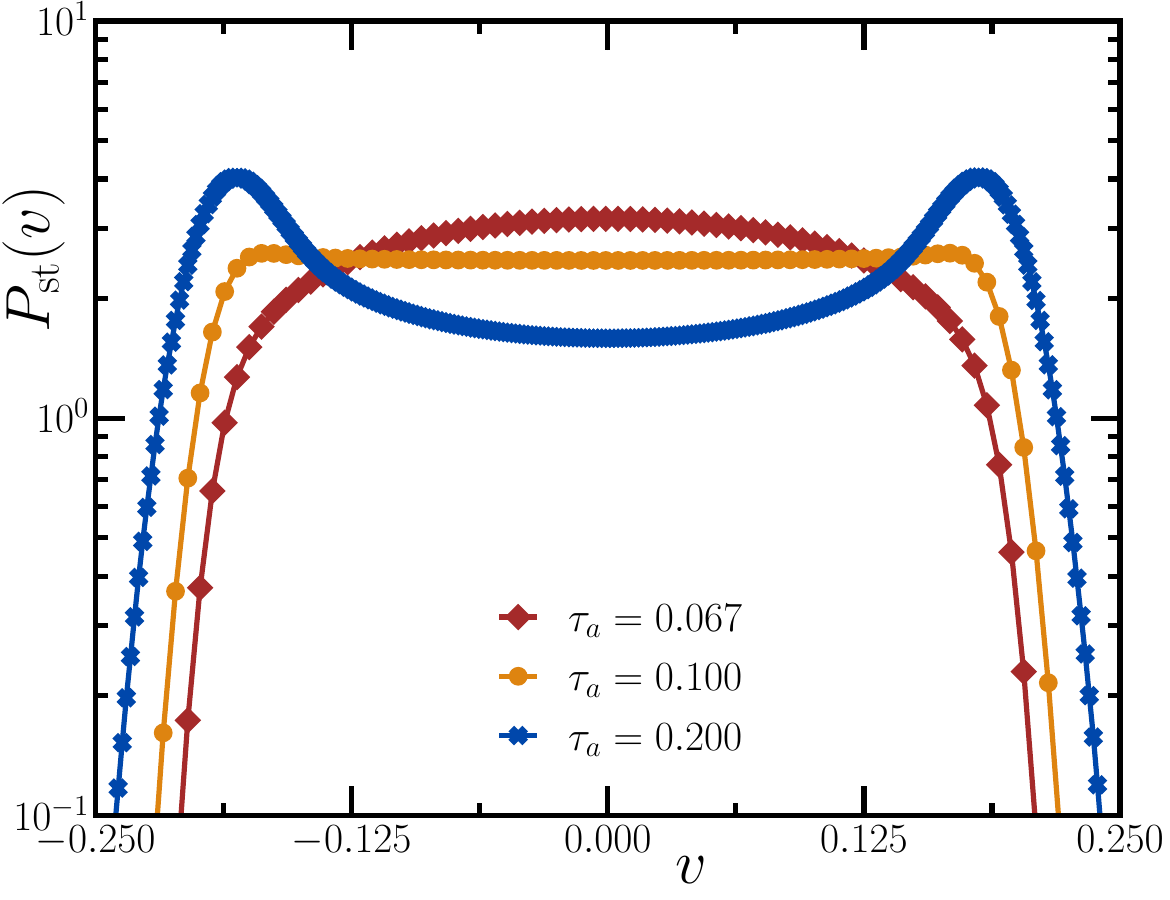}
    \caption{Overdamped moderately active and passive regime: Plot showing second velocity transition in the velocity distributions $P_{\text{st}}(v)$ by increasing activity $\tau_a$ beyond $\tau_2$. Here we have taken $m=0.5, \mu=0.5, a_{0}=1.0$ and $\gamma=5$.}
    \label{fig:od_approx_v}
\end{figure}
To understand the behaviour of the velocity distribution in the strongly active and the moderately active regimes, we use the same approximation introduced in Sec.~\ref{sec:OD_Pos}. Taking a time derivative of \eqref{eq:x_m_expd}, we get,
\bea
v=-\kappa z+\bar{v}_0\sigma+O(\tau_1)\label{eq:v_eff}.
\eea
where, $\kappa$ and $\nu$ are defined in \eref{eq:kappa_nu_def}. Keeping the leading order terms, i.e., ignoring the $O(1)$ terms, we can compute the stationary distribution of the velocity [see Appendix~\ref{App:sto_pxv} for details], quoted in \eref{eq:Pv_eff}.

Clearly, to this order, the velocity distribution $P_\text{st}(v)$ is bounded in the region $|v|<2\bar{v}_0$. Moreover, it predicts a shape transition in the velocity distribution at the critical activity $\tau_a^{\star}=\kappa^{-1}$ defined in \eref{eq:taus_m}. In the active regime $\tau_a>\tau_a^{\star}$, the distribution diverges at $v=0$ algebraically with exponent $1-1/(\tau_a\kappa)$, while in the passive regime $\tau_a<\tau_a^{\star}$ the distribution becomes bimodal with a minimum at $v=0$. Figure~\ref{fig:add_2RTP_Pv} compares the approximate velocity distribution given by \eref{eq:Pv_eff} with the data obtained from numerical simulations, which shows an excellent agreement between the two for a small value of $\tau_1$, i.e., in the strongly overdamped regime. 
\begin{figure}[t]
\centering
\includegraphics[width=8cm]{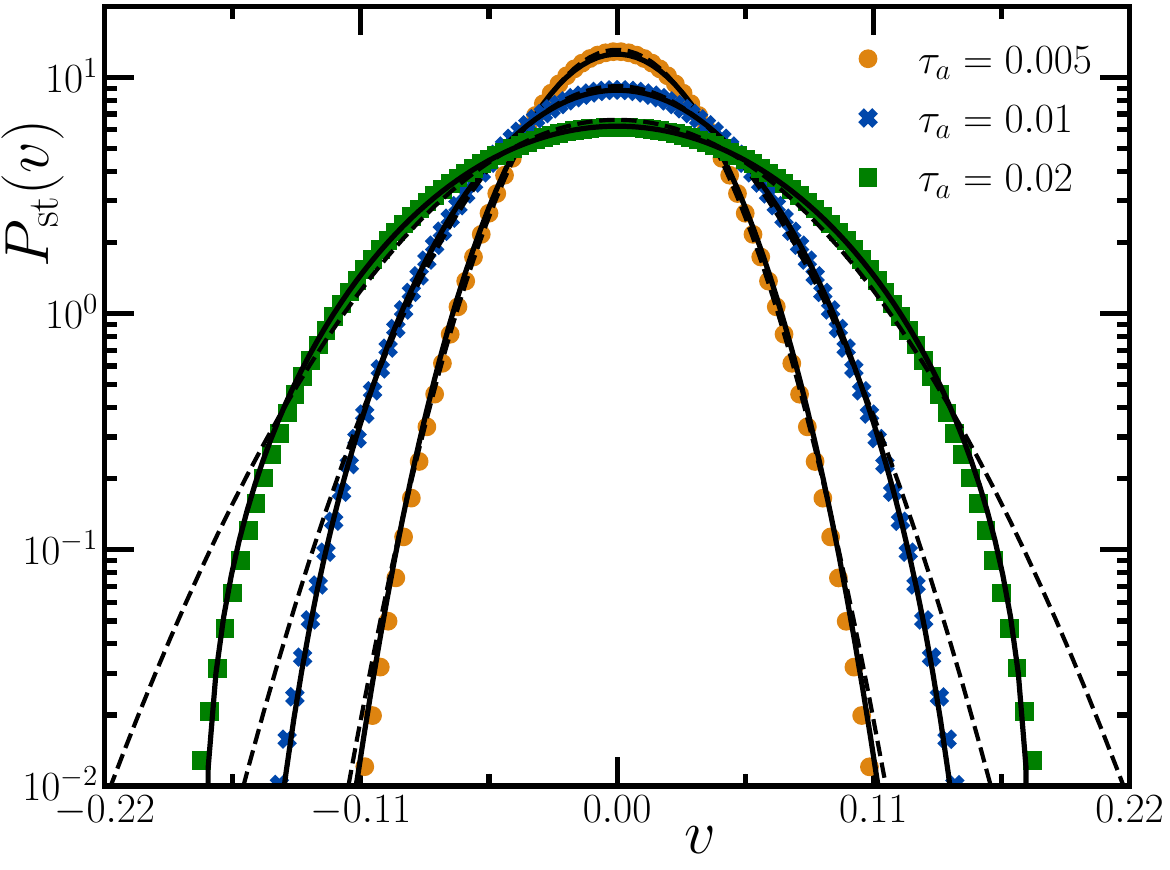}
\caption{Overdamped strongly passive regime: Plot showing the velocity
distribution $P_{\text{st}}(v)$ for different values of activity $\tau_a$. The symbols are for simulation results and the solid lines stand for the corrected Gaussian given by Eq. (\ref{eq:corr_Gauss_}). Here we have used $m=0.5,\mu=1.0$ and $\gamma=5$.}
\label{fig:dist_high_alp_v_od}
\end{figure}


As the activity is decreased further, the velocity distribution undergoes a second shape transition from the bimodal distribution to a dome-shaped distribution as shown in Fig.~\ref{fig:od_approx_v}. However, since the approximation \eref{eq:v_eff} ignores $O(1)$ terms, it works only in the strongly and moderately active regimes. We cannot use this approximation to obtain the velocity distribution in the passive regime, i.e., $\tau_a\ll\tau_1$. The behaviour of the velocity distribution in the passive regime can be obtained using the same argument as before---for $\tau_a\ll\tau_1$, the Langevin equation \eqref{eq:Leq2} reduces to that of a diffusive particle with an effective temperature $\tau_a a_0^2 / (2\gamma)$. Consequently, the typical velocity distribution approaches a Gaussian. As already discussed in Sec.~\ref{Pv_ud} for the underdamped scenario, the corrections to the Gaussian can be systematically computed by considering contributions from higher-order cumulants of the velocity. Figure~\ref{fig:dist_high_alp_v_od} compares the velocity distribution obtained using numerical simulations, with the analytical prediction \eqref{eq:corr_Gauss_} with $N=5$, which shows an excellent agreement. 

\section{Conclusion}\label{conc}

In this paper, we have studied the behaviour of an inertial run-and-tumble particle of mass $m$ in the presence of a harmonic trap of strength $\mu$. The active nature of the dynamics, characterized by the tumbling rate $\tau_a^{-1}$ gives rise to a nonequilibrium stationary state which is strikingly different than the equilibrium state of a passive inertial particle in a harmonic trap. We show that, the interplay of the three inherent time-scales present in the system, namely, the viscous time-scale $\tau_1 = m/\gamma$, trap time-scale $\tau_2 = \gamma /\mu$ and the activity time-scale $\tau_a$,  gives rise to a wide range of novel features. In particular, we find that, two qualitatively different scenarios emerge depending on whether $\tau_2 > 4\tau_1$ or not, i.e., whether the corresponding passive dynamics is overdamped or underdamped.  We characterize the behaviour of the IRTP by investigating the position and velocity distributions in these two scenarios. 

The marginal distributions in the underdamped scenario have certain unique features which are very distinct from that of ordinary RTP. Firstly, the distributions have a mass-dependent finite support which we compute explicitly. Secondly, both the position and the velocity distributions undergo shape transition from a multi-peaked structure in the strongly active regime ($\tau_1\ll\tau_a$) to a single-peaked one in the strongly passive regime ($\tau_a\ll\tau_1$). We characterize the multi-peaked distribution analytically using a simplified approximate dynamics. On the other hand, the marginal distributions in the passive regime result from fluctuations that are typically Gaussian. 

In the overdamped case, the presence of finite inertia also leads to various interesting features. The position distribution, in this case, is qualitatively similar to that of ordinary RTP, which has a mass-independent bound and undergoes a transition from a U-shaped distribution to a dome-shaped one as the activity is decreased. However, now the critical activity depends on the mass of the particle. The velocity distribution, which has a mass-dependent bound also undergoes a shape transition---from a distribution exhibiting algebraic divergence at $v=0$ in the strongly active regime to a bimodal distribution with minimum at $v=0$ in the moderately active regime. Interestingly, the effect of inertia is manifest by the presence of a second shape transition while moving from the moderately active regime to the strongly passive regime where the distribution eventually converges to a Gaussian. We analytically characterize these distributions in the strongly overdamped limit $\tau_1\ll\tau_2$.

Here we have investigated the marginal distributions of the IRTP in the presence of well-separated time scales. However, it remains an interesting open question to find the distributions when the time-scales are comparable. It would also be interesting to investigate the behaviour of an IRTP in higher dimensions, which is particularly relevant experimentally. Moreover, it would be intriguing to study the motion of an IRTP in the presence of a magnetic field to investigate questions like how the non-Gaussian nature of the active noise changes the behaviour compared to inertial AOUP \cite{muhsin2022inertial} and passive Brownian particles \cite{jimenez2006brownian}. Finally, IRTP provides a simple set-up to investigate how inertia affects the collective behaviour of active particles. 

\appendix
\section{Formal Solution of the Langevin Equation}\label{ap:solution_Langevin}
In this Appendix, we provide the formal solution of the Langevin \eref{eq:Leq2}. To solve these equations, which is a set of first-order linear coupled inhomogeneous equations, we first recast it in the form of a matrix equation,
\bea
\frac{dX}{dt}=MX+F, \label{eq: Leq_mat}
\eea
where,
\bea
X=\left[{x\atop v}\right],~M=\left[{0\atop -\frac \mu m}~{1\atop - \frac \gamma m}\right],~F(t)=\left[{0\atop \frac{a_{0}}{m}\sigma(t)}\right].
\eea
The formal solution to \eref{eq: Leq_mat} is given by,
\bea
X(t)=\Phi(t)X_{0}+\Phi(t)\int_{0}^{t}ds~\Phi^{-1}(s)F(s),\label{eq:Xformal}
\eea
where, $\Phi(t)=e^{Mt}$ and $X_{0}=[x_0~v_0]^\text{T}$ denotes the  initial condition on position and velocity. It is straightforward to compute $\Phi(t)$ explicitly, and it is given by,
\begin{align}
\Phi(t)=e^{-t/(2\tau_1)}
\begin{pmatrix}
\displaystyle {\cosh\frac{\omega t}{2\tau_1} \atop + \frac{1}{\omega}\sinh\frac{\omega t}{2\tau_1}} & \frac{2\tau_1}{\omega}\sinh\frac{\omega t}{2\tau_1} \\[1.5em] 
 -\frac{2}{\omega\tau_2}\sinh\frac{\omega t}{2\tau_1} & \displaystyle {\cosh\frac{\omega t}{2\tau_1} \atop  -\frac{1}{\omega}\sinh\frac{\omega t}{2\tau_1}}
\end{pmatrix},
\end{align}
where, $\omega=\sqrt{1-4 \tau_1 / \tau_2}$. Moreover,
\begin{align}
\Phi^{-1}(s)F(s)=\frac{a_0\sigma(s)}{m}e^{s/(2\tau_1)}\left[\begin{array}{c}
-\frac{2\tau_1}{\omega}\sinh\frac{\omega s}{2\tau_1}\\
\cosh\frac{\omega s}{2\tau_1}+\frac{1}{\omega}\sinh\frac{\omega s}{2\tau_1}
\end{array}\right].
\end{align}
Substituting the above matrix in \eref{eq:Xformal}, we get the general solution,
\begin{widetext}
\begin{align}
x(t) & =e^{-\frac{t}{2\tau_1}}\left[x_{0}(\cosh\frac{\omega t}{2\tau_1}+\frac{1}{\omega}\sinh\frac{\omega t}{2\tau_1})+\frac{2\tau_1 v_{0}}{\omega}\sinh\frac{\omega t}{2\tau_1}\right]+\frac{2\bar{v}_{0}}{\omega}\int_{0}^{t}ds~\sigma(s)e^{-\frac{1}{2\tau_1}(t-s)}\sinh\left[\frac{\omega}{2\tau_1}(t-s)\right],\label{eq:x_formal}\\
v(t)&=e^{-\frac{t}{2\tau_1}}\left[-\frac{2 x_{0}}{\tau_2 \omega}\sinh\frac{\omega t}{2\tau_1}+v_{0}(\cosh\frac{\omega t}{2\tau_1}-\frac{1}{\omega}\sinh\frac{\omega t}{2\tau_1})\right]\cr
&\hspace{120pt} +\frac{\bar{v}_{0}}{\tau_1}\int_{0}^{t}ds~\sigma(s)e^{-\frac{1}{2\tau_1}(t-s)}\Bigg[\cosh\Big[\frac{\omega}{2\tau_1}(t-s)\Big]-\frac{1}{\omega}\sinh\Big[\frac{\omega}{2\tau_1}(t-s)\Big]\Bigg]\label{eq:v_formal},
\end{align}
\end{widetext}
where $\bar{v}_0=a_0/\gamma$ is defined in \eqref{eq:v0_mu_def}. For the overdamped case, $4\tau_1\le\tau_2$, and using initial conditions, $x=0$ and $v=0$ at $t=0$, we recover \erefs{eq:xt_od}-\eqref{eq:vt_od} by identifying $\omega=\lambda$. On the other hand, for the underdamped case, $4\tau_1>\tau_2$, we substitute $\omega=i\Lambda$ in \erefs{eq:x_formal}-\eqref{eq:v_formal}, which gives \erefs{eq:xt_ud}-\eqref{eq:vt_ud} using the same set initial conditions, $x=0$ and $v=0$ at $t=0$.

\section{Computation of moments from Langevin equation}\label{appB}
In this section, we explicitly compute the exact time-dependent second-order moments and correlations of position and velocity, using the Langevin \eref{eq:Leq2}. To this end, we require the autocorrelation function for the dichotomous noise $\sigma(t)$ \cite{haunggi1994colored}, which reads,
\begin{align}
\la \sigma(t_1)\sigma(t_2) \ra = e^{-2|t_1 - t_2|/\tau_a}\label{eq:autocorr}    
\end{align}
Using the initial condition $x_0=v_0=0$ for simplicity, the correlations can then be explicitly computed using \erefs{eq:x_formal} and \ref{eq:v_formal}). This yields, for the overdamped case,
\begin{widetext}
\begin{align}
\la x^{2}(t) \ra &= \frac{\bar{v}_0^2 \tau_2}{\lambda^2\Big[\big(\frac{4\tau_1}{\tau_a^2}+\frac{1}{\tau_2}\big)^2-\frac{4}{\tau_a^2}\Big]}\Bigg[e^{-t/\tau_1}\Big[\frac{2}{\tau_a}\big(\frac{2\tau_1}{\tau_a}+1\big)+\frac{1}{\tau_2}\Big]\Big[\frac{2\tau_1}{\tau_2}\big(\frac{4\tau_1}{\tau_a}-1\big)+\big(1-\frac{2\tau_1}{\tau_a}-\frac{2\tau_1}{\tau_2}\big)\cosh\frac{\lambda t}{\tau_1}-\big(\frac{2\tau_1}{\tau_a}-1\big)\cr
&\lambda\sinh\frac{\lambda t}{\tau_1}\Big]+\lambda^2\Big(\frac{2\tau_1}{\tau_a}+1\Big)\Big(\frac{4\tau_1}{\tau_a^2}-\frac{2}{\tau_a}+\frac{1}{\tau_2}\Big)-\frac{2\lambda}{\tau_2}e^{-t(4\tau_1/\tau_a+1)/(2\tau_1)}\Big[\lambda\cosh\frac{\lambda t}{2\tau_1}+\big(\frac{4\tau_1}{\tau_a}+1\big)\sinh\frac{\lambda t}{2\tau_1}\Big]\Bigg],\label{eq:x2t_full_od}
\\
\la v^{2}(t) \ra &= \frac{8\bar{v}_0^2\tau_1/\tau_2}{\lambda ^2\Big[\lambda ^2-\big(\frac{4\tau_1}{\tau_a}+1\big)^2\Big]\Big[\lambda ^2-\big(\frac{4\tau_1}{\tau_a}-1\big)^2\Big]}\Bigg[16\lambda\frac{\tau_1}{\tau_a}e^{-t(4\tau_1/\tau_a+1)/(2\tau_1)}\Big[\frac{\tau_2}{\tau_a}\lambda\cosh\frac{\lambda t}{2\tau_1}-\big(1+\frac{\tau_2}{\tau_a}\big) \sinh\frac{\lambda t}{2\tau_1}\Big]\cr
&-\frac{\tau_2}{\tau_a}\lambda ^2\Big[\lambda ^2-\big(\frac{4\tau_1}{\tau_2}-1\big)^2\Big]-e^{-t/\tau_1 }\Big[\lambda ^2-\big(1+\frac{4\tau_1}{\tau_a}\big)^2\Big]\Big[\frac{4\tau_1}{\tau_a}-1+\big(1-\frac{\tau_2}{\tau_a}\big)\cosh\frac{\lambda t}{\tau_1}+\frac{\tau_2}{\tau_a}\lambda \sinh\frac{\lambda t}{\tau_1}\Big]\Bigg],\label{eq:v2t_full_od}
\\
\la x(t) v(t) \ra &= \frac{8\bar{v}_0^2\tau_1 \sinh \frac{\lambda t}{2\tau_1}}{\lambda ^2\big[\lambda ^2-(\frac{4\tau_1}{\tau_a}-1)^2\big]}\Bigg[e^{-t/\tau_1}\Big[\lambda \cosh\frac{\lambda t}{2\tau_1}-\big(\frac{4\tau_1}{\tau_a} - 1\big)\sinh \frac{\lambda  t}{2\tau_1}\Big]-\lambda  e^{-t(4\tau_1/\tau_a+1)/(2\tau_1)}\Bigg],\label{eq:xvt_full_od}
\end{align}
\end{widetext}
where $\lambda$ is defined in \eref{eq:w_def}. Similarly, in the underdamped case, the correlations take the following form,
\begin{widetext}
\begin{align}
\la x^{2}(t) \ra &= \frac{\bar{v}_0^2 \tau_2}{\Lambda^2\Big[\big(\frac{4\tau_1}{\tau_a^2}+\frac{1}{\tau_2}\big)^2-\frac{4}{\tau_a^2}\Big]}\Bigg[-e^{-t/\tau_1}\Big[\frac{2}{\tau_a}\big(\frac{2\tau_1}{\tau_a}+1\big)+\frac{1}{\tau_2}\Big]\Big[\frac{2\tau_1}{\tau_2}\big(\frac{4\tau_1}{\tau_a}-1\big)+\big(1-\frac{2\tau_1}{\tau_a}-\frac{2\tau_1}{\tau_2}\big)\cos\frac{\Lambda t}{\tau_1}+\big(\frac{2\tau_1}{\tau_a}-1\big)\cr
&\Lambda\sin\frac{\Lambda t}{\tau_1}\Big]+\Lambda^2\Big(\frac{2\tau_1}{\tau_a}+1\Big)\Big(\frac{4\tau_1}{\tau_a^2}-\frac{2}{\tau_a}+\frac{1}{\tau_2}\Big)-\frac{2\Lambda}{\tau_2}e^{-t(4\tau_1/\tau_a+1)/(2\tau_1)}\Big[\Lambda\cos\frac{\Lambda t}{2\tau_1}+\big(\frac{4\tau_1}{\tau_a}+1\big)\sin\frac{\Lambda t}{2\tau_1}\Big]\Bigg],\label{eq:x2t_full_ud}
\\
\la v^{2}(t) \ra &= \frac{8\bar{v}_0^2\tau_1/\tau_2}{\Lambda ^2\Big[\Lambda ^2+\big(\frac{4\tau_1}{\tau_a}+1\big)^2\Big]\Big[\Lambda ^2+\big(\frac{4\tau_1}{\tau_a}-1\big)^2\Big]}\Bigg[16\Lambda\frac{\tau_1}{\tau_a}e^{-t(4\tau_1/\tau_a+1)/(2\tau_1)}\Big[\frac{\tau_2}{\tau_a}\Lambda\cos\frac{\Lambda t}{2\tau_1}-\big(1+\frac{\tau_2}{\tau_a}\big) \sin\frac{\Lambda t}{2\tau_1}\Big]\cr
&+\frac{\tau_2}{\tau_a}\Lambda ^2\Big[\Lambda ^2+\big(\frac{4\tau_1}{\tau_2}-1\big)^2\Big]-e^{-t/\tau_1 }\Big[\Lambda ^2+\big(1+\frac{4\tau_1}{\tau_a}\big)^2\Big]\Big[\frac{4\tau_1}{\tau_a}-1+\big(1-\frac{\tau_2}{\tau_a}\big)\cos\frac{\Lambda t}{\tau_1}-\frac{\tau_2}{\tau_a}\Lambda \sin\frac{\Lambda t}{\tau_1}\Big]\Bigg],\label{eq:v2t_full_ud}
\\
\la x(t) v(t) \ra &= \frac{8\bar{v}_0^2\tau_1 \sin \frac{\Lambda t}{2\tau_1}}{\Lambda^2\big[\Lambda^2+(\frac{4\tau_1}{\tau_a}-1)^2\big]}\Bigg[e^{-t/\tau_1}\Big[\big(\frac{4\tau_1}{\tau_a} - 1\big)\sin\frac{\Lambda t}{2\tau_1} - \Lambda\cos\frac{\Lambda t}{2\tau_1}\Big]+\Lambda e^{-t(4\tau_1/\tau_a+1)/(2\tau_1)}\Bigg].\label{eq:xvt_full_ud}
\end{align}
\end{widetext}
where $\Lambda$ is defined in \eref{eq:Lamb_def}.
\section{Calcuation of stationary state distribution bound}\label{appC}

In this section, we explicitly compute the bounds of the stationary state distribution. As discussed in Sec.~\ref{ud_case}, in the underdamped case, this bound is obtained from the trajectory which maximizes the displacement of the particle --- when tumbling events $\sigma\to-\sigma$ coincide with the velocity $v$ changing its sign. To construct such a trajectory, we first compute the position of a particle at the turning points of its underdamped motion. To this end, we evaluate the integrals in \erefs{eq:x_formal} and \eqref{eq:v_formal} with the particle starting at rest with some initial position $x_{0}$ and for a fixed propulsion direction $\sigma$, 
\begin{align}
    x(t) &= \frac{a_0}{\mu}\sigma+(x_0-\frac{a_0}{\mu}\sigma)e^{-t/(2\tau_1)}\Big[\cos\frac{\Lambda t}{2\tau_1}+\frac{1}{\Lambda}\sin\frac{\Lambda t}{2\tau_1}\Big],\label{eq:xt_det}\\
    v(t) &= \frac{2 e^{-t/(2\tau_1)}}{\Lambda \tau_2}\big(\frac{a_0}{\mu}\sigma-x_0\big)\sin\frac{\Lambda t}{2\tau_1},\label{eq:vt_det}
\end{align}
where $\Lambda$ is defined in \eref{eq:Lamb_def}. Then the time instants of the turning points are computed by solving for the roots of the equation $v(t)=0$,
\begin{figure}
\centering
\includegraphics[width=8.0 cm]{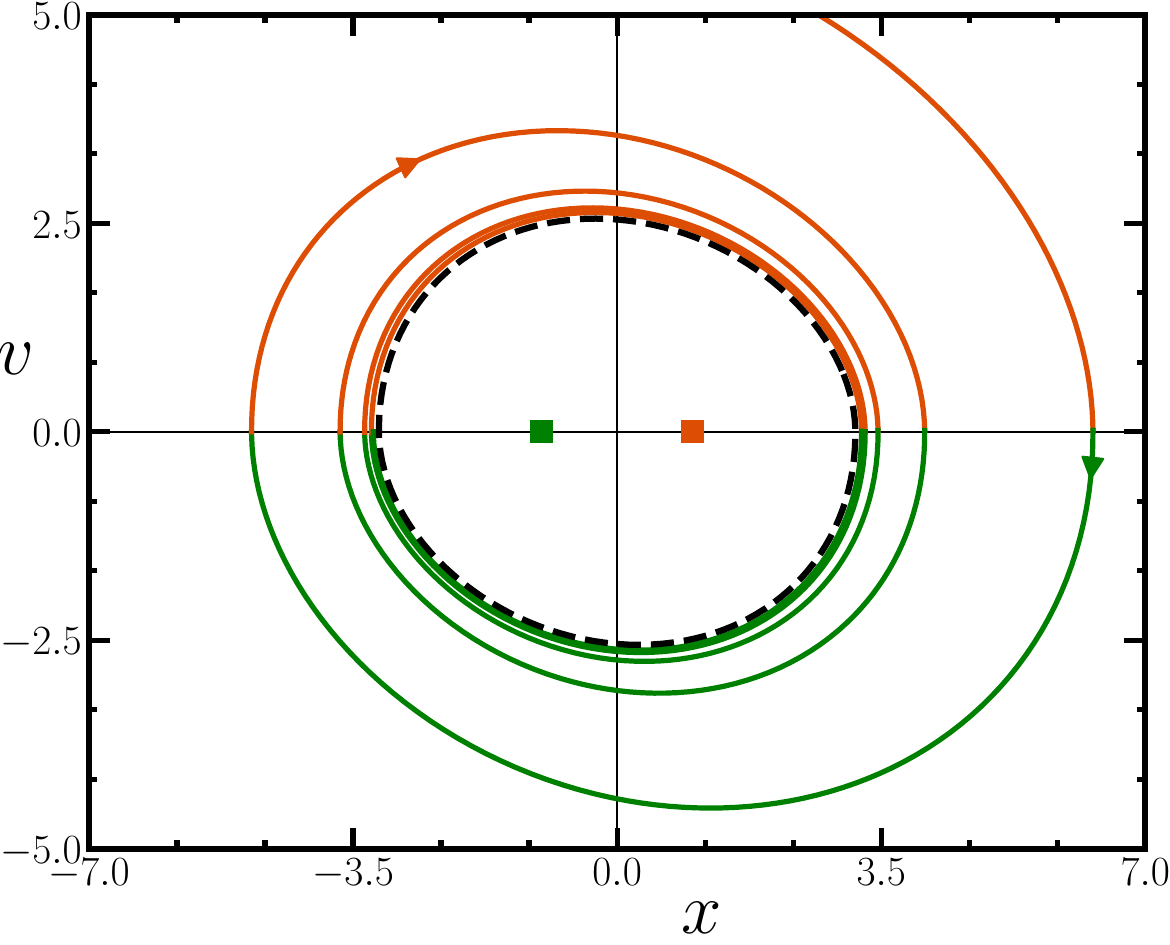}
\caption{Underdamped scenario: Typical displacement maximizing trajectory of IRTP with initial position ($x_0=0$) and velocity ($v_0=1$) chosen outside the supported region in the $x-v$ plane. The particle eventually settles into a limit cycle (black dashed line) described by the trajectories \eref{eq:lcyc_1}-\eqref{eq:lcyc_2}, which gives the bound of joint stationary distribution $P_{\text {st}}(x,v)$. Here we have used $\gamma=0.5,\mu=1.0,a_0=1.0$ and $m=1.5$.}
\label{fig:bound_traj_ud_}
\end{figure}
\bea
t_{n}=\frac{2\pi n\tau_1}{\Lambda},\quad n=1,2,3,\ldots
\eea
Consequently, the position of the $n$-th turning point is given by substituting $t_{n}$ in \eref{eq:xt_det},
\bea
x_{n}=\frac{a_{0}}{\mu}\sigma + (-1)^{n}e^{-n\pi/\Lambda}(x_{0} - \frac{a_{0}}{\mu}\sigma). \label{eq:tp_ud}
\eea
By definition, the tumbling event occurs at the first turning point $x_{1}$. Consequently, we have the underdamped particle now starting at rest with initial position $x_{1}$ and with self-propulsion direction $-\sigma$. The next tumbling event again occurs at the first turning point of the next trajectory, and so on --- iteratively generating the full trajectory. 
    
Let $x_{\ell}$ denote the particle position at the $\ell$-th tumbling event $\sigma_{\ell}=-\sigma_{\ell-1}$ in the trajectory. Considering the iterative nature of the trajectory, we obtain a recursive relation for $x_{\ell}$, by substituting $x_{\ell-1}$ as the initial position and $\sigma_{\ell - 1}$ as the propulsion direction in \eref{eq:tp_ud},
\bea
x_{\ell}=c x_{\ell-1}+ \frac{a_0}{\mu} (1 - c)\sigma_{\ell-1},\label{eq:rec_rel}
\eea
where $c=-e^{-\pi / \Lambda}$ and $\Lambda$ is defined in \eref{eq:Lamb_def}. Depending on the sign of $|x_{\ell}|-|x_{\ell-1}|$, the particle distance from the origin can either increase or decrease at successive tumbling events. The two scenarios for $|x_0|<a_0/\mu$ and $|x_0|>a_0/\mu$ are illustrated in Figs.~\ref{fig:bound_traj_ud} and \ref{fig:bound_traj_ud_} respectively. Consequently, the bound can be computed from the asymptotic value of $x_{\ell}$ in the limit $\ell\to\infty$. Assuming that the particle starts at rest with some initial position $x_0$, and propulsion direction $\sigma_0$, the position of the $\ell$-th turning point can be computed iteratively, using \eref{eq:rec_rel}, as quoted in \eref{eq:rec_sol}. Then, the maximum distance that the particle can reach is given by the limiting value $\lim_{\ell\to\infty}x_{\ell}$, which we identify as the bound of the stationary position distribution quoted in \eqref{eq:x_b}, 

Alternately, the bound can also be obtained using the condition, $|x_{l}|=|x_{l-1}|$, which gives the limit cycle describing the boundary of the joint distribution $P_{\text{st}}(x,v)$. Then by setting $x_{\ell}=-x_{\ell-1}$ in \eref{eq:rec_rel}, we obtain the bound for the position distribution explicitly. 

Similarly, the bound for the velocity distribution is given by the maximum attained velocity in this limit cycle, i.e. when the particle acceleration changes sign for the first time. To locate this velocity turning point,  we first compute the time instants of these turning points by solving for the roots of the equation $dv/dt=0$, using \eref{eq:vt_det}, which gives,
\bea
t_{n}=\frac{2\tau_1}{\Lambda}(\tan^{-1}\Lambda+n\pi),\quad n=0,1,2,\ldots.
\eea
Then, the $n$-th turning point in the velocity is computed by evaluating the velocity $v(t_n)$ at time $t_{n}$
\begin{align}
v_{n} = \frac{(-1)^{n}}{\sqrt{m/\mu}}\Big(\frac{a_0}{\mu}\sigma-x_0\Big)e^{-\frac{1}{\Lambda} \big[n\pi + \tan^{-1}\Lambda\big]}\label{eq:tp_ud_}    
\end{align}
It is then straightforward to calculate the velocity bound from the first turning point $(n=0)$ for the particle starting from an initial position $x_0=-\sigma X_\text{UD}$, which gives us the expression quoted in \eref{eq:v_b}. 

Lastly, we note that these turning points also allow us to compute the position of the peaks in the position and velocity distributions. By substituting $x_0=\pm a_0/\mu$ in \erefs{eq:tp_ud} and \eqref{eq:tp_ud}, we arrive at the expressions quoted in \erefs{eq: peak_p} and \eqref{eq:ud_activ_vd_peaks} respectively.

We now turn to the overdamped scenario. As already discussed in Sec.~\ref{sec:OD}, in the overdamped case, the displacement of the particle is maximum when no tumbling event takes place. This is easily seen by examining the integrand in the formal solution for $x(t)$, quoted in \eref{eq:x_formal}. Apart from $\sigma(s)$, the integrand is a purely positive quantity, i.e., in a given time interval the integral evaluates to the maximum value when no tumbling event takes place. Performing the integral in \erefs{eq:x_formal} and \eqref{eq:v_formal} with particle starting at with position $x_0$ and with a fixed propulsion direction $\sigma$, we get for the overdamped case,
\begin{align}
    x(t) &=\frac{a_0}{\mu}\sigma+(x_0-\frac{a_0}{\mu}\sigma)e^{-t/(2\tau_1)}\Big[\cosh\frac{\lambda t}{2\tau_1}+\frac{1}{\lambda}\sin\frac{\lambda t}{2\tau_1}\Big].\label{eq:xt_det_}\\
    v(t) &=\frac{2 e^{-t/(2\tau_1)}}{\lambda \tau_2}\big(\frac{a_0}{\mu}\sigma-x_0\big)\sinh\frac{\lambda t}{2\tau_1},\label{eq:vt_det_}
\end{align}
where $\lambda$ is defined \eref{eq:w_def}. Taking the limit, $t\to\infty$ in \eref{eq:xt_det_}, gives the maximum displacement of the particle for a given propulsion direction, which we identify as the bound of the position distribution, $X_{\text{OD}}$, in the overdamped case, quoted in \eref{eq:x_b_}. Consequently, the joint distribution must be bounded by the trajectories connecting the point $(\pm X_\text{OD},0)$, allowing us to compute the bound on the velocity distribution. 

In particular, the velocity extremum between the points $(\pm X_\text{OD}, 0)$ marks the velocity bound in this case. We compute this bound explicitly by first solving for the equation $dv/dt=0$, using \eref{eq:vt_det_}, which gives,
\bea
t=\frac{2\tau_1}{\lambda}\cosh^{-1}\frac{\gamma}{2\sqrt{m \mu}},\label{eq:tv_OD}
\eea
and by computing the velocity at this time instant with initial position $x_0=-\sigma X_\text{OD}$ in \eref{eq:vt_det_}, which gives the expression quoted in \eref{eq:v_b_}.

\section{Effective models in the strongly overdamped limit}\label{App:sto_pxv}
In this Appendix, we explicitly compute the marginal position distribution in the strongly overdamped limit. We start from \eref{eq:xz_coup_2}, the Langevin equations for the variables $x(t)=x_{1}(t)-x_{2}(t)$ and $z(t)=x_{1}(t)+x_{2}(t)$,
\begin{align}
    \dot{x}(t)=&-k_+ x(t) +k_- z(t), \label{eq:xz_coupl_1}\\ 
    \dot{z}(t)=&-k_+ z(t) + k_- x(t) + 2\bar{a}_0\sigma(t).\label{eq:xz_coupl_2}
\end{align}
where, $k_\pm$ are defined in \eref{eq:kpm_def}. In the strongly overdamped limit, $\tau_1\ll\tau_2$, we treat $\tau_1$ as the small parameter and expand $x(t)$ in orders of $\tau_1$ using \eref{eq:xz_coupl_1}, to get,
\bea
    x=z-2\Big(\dot{x}+\frac{z}{\tau_2}\Big)\tau_1-2\Big(\frac{\tau_1}{\tau_2}\Big)^2\dot{z}+O(\tau_1^{3}).\label{eq:x_expd_eps}
\eea
By taking time derivative of the above \eref{eq:x_expd_eps}, we get,
\bea
    \dot{x}=\dot{z}-2\Big(\ddot{x}+\frac{z}{\tau_2}\Big)\tau_1-2\Big(\frac{\tau_1}{\tau_2}\Big)^2\ddot{z}+O(\tau_1^{3}).\label{eq:xdot_expd_eps}
\eea
Substituting \eref{eq:xdot_expd_eps} in \eref{eq:x_expd_eps} and keeping up to the leading order term in $\tau_1$, we get,
\begin{align}
x\approx \Big(1-2\frac{\tau_1}{\tau_2}\Big)z -2\tau_1\dot{z}.\label{eq:x_expd_z}
\end{align}
Substituting  $x(t)$ \eref{eq:x_expd_z} in \eqref{eq:xz_coupl_2} we get an effective equation for $z(t)$, quoted in \eref{eq:zeff}.
Clearly, $z(t)$ is a fully overdamped RTP moving in a harmonic trap of stiffness $\kappa$ with propulsion speed $\nu$. The joint distribution $P_{\sigma}(z)$ is known explicitly \cite{dhar2019run}, and is given by,
\begin{align}
P_{\sigma}(z)=\frac{4^{-\alpha}\kappa/\nu}{B(\alpha,\alpha)}\frac{[1-(\frac{\kappa z}{\nu})^2]^{\alpha}}{\Big(1-\sigma\frac{\kappa z}{\nu}\Big)},\label{eq:Psigmaz}    
\end{align}
where $\alpha=1/(\tau_a\kappa)$, and is bounded in the region $-\nu/\kappa<z<\nu/\kappa$. The distribution of $x$ can be obtained using \eqref{eq:x_expd_z}, \eqref{eq:zeff} and \eqref{eq:Psigmaz}. Substituting \eref{eq:zeff} in \eref{eq:x_expd_z} and keeping upto the leading order term, we get,
\bea
    x(t)=az(t)-b\sigma(t),
\eea
with,
\bea
a=\frac{\lambda  \tau_2}{\tau_2-2\tau_1}~~\text{and}~~b=2 \bar{v}_0\tau_1 a.
\eea
Thus the stationary distribution of $x$ is given by, 
\begin{align}
    P_\text{st}(x)=\sum_{\sigma}\intop_{-\nu/\kappa}^{\nu/\kappa}dz~\delta(x-az+b\sigma)P_\sigma(z).\label{eq:Px_eff_formal}
\end{align}
where, $P_{\sigma}(z)$ is defined in \eref{eq:Psigmaz} and $\delta(u)$ is the delta function. Performing the integral over $z$ we get,
\begin{align}
    P_\text{st}(x)=\frac{1}{a}\sum_{\sigma=\pm 1}P_{\sigma}\Big(\frac{x}{a}+\frac{b}{a}\sigma\Big)\Theta\Big(\frac{\nu}{\kappa}a+b\sigma+x\Big)\cr\Theta\Big(\frac{\nu}{\kappa}a-b\sigma-x\Big).
\end{align}
To order $O(\tau_1)$, $a=1$ and we arrive at the expression quoted in \eref{eq:Px_eff}. 

By definition, the velocity $v$ is given by the time derivative of \eref{eq:x_expd_z}. To the leading order in $\tau$, the Langevin equation for the velocity $v$ is then given by,
\bea
v=\dot{z}+O(\tau_1)
\eea
Consequently, the stationary velocity distribution is given by,
\bea
P_{\text{st}}(v)=\sum_{\sigma=\pm 1}\intop_{-\kappa /\nu}^{\kappa /\nu}dz~\delta(v+\kappa z-\nu\sigma)P_{\sigma}(z).\label{eq:Pv_eff_formal}
\eea
where, $P_{\sigma}(z)$ is defined in \eref{eq:Psigmaz}. Performing the integral over $z$ we obtain \eref{eq:Pv_eff}.

\bibliography{cite}

\end{document}